\documentclass[pra,twocolumn,floatfix,superscriptaddress]{revtex4-1}
\usepackage{dcolumn,amsmath}
\usepackage{graphicx}
\usepackage{bm}
\usepackage{hyperref}

\usepackage[utf8]{inputenc}
\usepackage[T2A]{fontenc}

\usepackage{float}

\usepackage{hyperref}
\hypersetup{colorlinks=true, citecolor=blue, urlcolor=blue, linkcolor=blue}

\usepackage{accents}
\newcommand{\dbtilde}[1]{\accentset{\approx}{#1}}

\usepackage{xcolor}

\begin{document}

\title{Revisited nuclear magnetic dipole and electric quadrupole moments of polonium isotopes}

\date{November 1, 2023}

\begin{abstract}

We revisited the electronic structure parameters used to interpret the hyperfine structure of neutral polonium. 
We used a computational scheme that treats relativistic and high-order electronic correlation effects within the coupled cluster with single, double, triple and perturbative quadruple excitations CCSDT(Q) method, as well as estimate the  contribution of quantum electrodynamics and finite nuclear size effects. A systematic study of the uncertainty is carried out. This allowed us to obtain significantly refined values for the nuclear magnetic dipole and electric quadrupole moments of a wide range of odd-mass polonium isotopes. For $^{205}$Po and $^{207}$Po we extracted both the magnetic moment and the nuclear magnetization distribution parameter in a nuclear model-independent way. To assess the accuracy of the calculations, we also computed the ionization potential (IP), excitation energies (EE) of the $6p^4~{}^1D_2$ and $6p^3 7s^1~{}^5S_2$ electronic states and the electronic $g_J$ factor in the same theoretical framework. A good agreement of the theory and experiment for IP, EEs and $g_J$ confirms the reliability of the computational scheme and uncertainty estimation for the Po electromagnetic moments. We identify the $6p^4~{}^1D_2$ electronic level as a potentially promising state for further studies of the nuclear moments of polonium isotopes.

\end{abstract}

\author{Leonid V.\ Skripnikov}
\email{skripnikov\_lv@pnpi.nrcki.ru,\\ leonidos239@gmail.com}
\homepage{http://www.qchem.pnpi.spb.ru}
\affiliation{Petersburg Nuclear Physics Institute named by B.P. Konstantinov of National Research Centre
``Kurchatov Institute'', Gatchina, Leningrad District 188300, Russia}
\affiliation{Saint Petersburg State University, 7/9 Universitetskaya nab., St. Petersburg, 199034 Russia}
\author{Anatoly E.\ Barzakh}
\affiliation{Petersburg Nuclear Physics Institute named by B.P. Konstantinov of National Research Centre
``Kurchatov Institute'', Gatchina, Leningrad District 188300, Russia}

\maketitle

\section{Introduction}

The investigation of nuclear magnetic dipole and electric quadrupole moments is of great significance in various fields of physics. These values serve as a means to test the predictions of nuclear theory models~\cite{Yang2022_exotic}. For example, electromagnetic moments (especially in the long isotopic chains) prove to be the key observables to adjust coupling constants in the time-odd mean-field sector of the energy-density functional~\cite{Sassarini:2022,NucDFT:2023}. Accurate values of the nuclear magnetic moments are required to test predictions of bound-state quantum electrodynamics using highly-charged ions~\cite{Shabaev:01a,Skripnikov:18a,Nortershauser:2019,Skripnikov:2022,Skripnikov:2020a}. The values of nuclear magnetic moments are sensitive to the single-particle degrees of freedom whereas the nuclear quadrupole moments provide information on nuclear deformation and can be regarded
 as an indicator of collective effects~\cite{Yang2022_exotic}. They are also important in chemical spectroscopy, as spectroscopic quadrupole splittings act as a gauge of the electron distribution~\cite{Pekka:2008}.

The calculation of hyperfine structure (HFS) constants serves as a standard test for the accuracy of electronic wave functions~\cite{Kozlov:97c,Safronova:18,Porsev:2009,ginges2017ground,Skripnikov:2020b,GFreview,KL95,Quiney:98,Titov:06amin,Skripnikov:15b,Skripnikov:15a,Sunaga:16,Fleig:17,Borschevsky:2020,Skripnikov:2020e} and is crucial for probing electronic structure methods, which are used in the interpretation of experiments aimed at searching for violations of symmetries in fundamental interactions in molecules and atoms, such as the electron electric dipole moments, nuclear anapole moment, etc. 

The analysis of the Po chain is important to study the effects of the shape coexistence and deformation, and to compare the isotopic trends of nuclear moments and radii with other isotopic chains in the lead region (see~\cite{Seliverstov:2014_Po} and references therein). 
The HFS parameters and other atomic observables of polonium has been widely studied experimentally. The atomic-beam magnetic resonance approach was used to study hyperfine interactions induced by the magnetic dipole and electric quadrupole nuclear moments of $^{205}$Po and $^{207}$Po with nuclear spin $I=5/2$ in the ground electronic state~\cite{olsmats1961hyperfine}. Hyperfine structure parameters of the longest-lived $^{209}$Po were measured in Ref.~\cite{Kowalewska:1991} using laser-induced fluorescence spectroscopy in an atomic beam. In Refs.~\cite{Seliverstov:2014_Po,Fink:2015}, 
hyperfine parameters of short-lived isotopes were determined. The electronic excitation energies (EE), ionization potential (IP) and the ground state electronic $g_J$ factor in Po were measured in  Refs.~\cite{Finkelnburg:1950,Charles:55,Charles:66,Raeder:2019,Po_IPexp:2019,axensten1961nuclear}.

The magnetic moments of $^{205,207}$Po were determined by the nuclear magnetic resonance on oriented nuclei (NMR), 0.760(55) $\mu_\mathit{N}$ and 0.793(55) $\mu_\mathit{N}$, respectively~\cite{herzog1983nuclear}. These values were used to deduce magnetic moments for other Po isotopes by scaling relation taking into account their measured HFS constants ~\cite{Kowalewska:1991,Seliverstov:2014_Po,Fink:2015}. However, the magnetic moments uncertainties stemming from the uncertainties of the reference values in the majority of cases prove to be larger than the statistical experimental errors by a factor of 3-7~\cite{Seliverstov:2014_Po}. 
To extract the spectroscopic quadrupole moment $Q_S$ from the measured HFS constant, semiempirical atomic calculations were implemented~\cite{olsmats1961hyperfine}. The result, $Q_S$($^{207}$Po)=$0.28~$b, was obtained and only the lower limit of its uncertainty (10\%) was indicated. This value with additional 20\% correction due to Sternheimer effect, was used in~\cite{Seliverstov:2014_Po,Fink:2015} as the reference for the scaling relation for the extraction of $Q_S$ in other Po isotopes with the measured HFS constants. Herewith, the theoretical uncertainty of these $Q_S$ values was arbitrarily set to 10\%. To get an idea of the possible accuracy of the reference $Q_S$ value one can compare $\mu(^{207}{\rm Po})=0.27~\mu_\mathit{N}$ derived in~\cite{olsmats1961hyperfine} within the framework of the same semiempirical approach with the directly measured value $\mu_{\rm expt}(^{207}{\rm Po})=0.793(55)~\mu_\mathit{N}$~\cite{herzog1983nuclear}. Thus, the theoretical uncertainty of this type of calculations can be significantly larger than 10\%. It can be concluded that currently we do not have reliable values of $Q_S$ for polonium isotopes and their uncertainties are not firmly established.

The aim of the present paper is to remove the existent indeterminacy in the interpretation of the HFS data for Po isotopes. By employing relativistic coupled cluster theory, we determine the values of the nuclear magnetic dipole and electric quadrupole moments of Po isotopes. In the case of the magnetic dipole moment, it is especially important to consider not only pure electronic effects but also the effects of the extended nucleus. Namely, the Breit-Rosenthal (BR) correction~\cite{rosenthal1932isotope, crawford1949mf}, which accounts for the finite charge distribution across the nucleus, and the Bohr-Weisskopf (BW) correction~\cite{bohr1950influence, bohr1951bohr, sliv1951uchet}, which considers the finite nuclear magnetization distribution should be taken into account. Although the treatment of the BR effect is quite simple, the consideration of the BW effect is more complex. Generally, this is because the calculation of the nuclear magnetization distribution requires a many-body nuclear structure treatment. In our analysis, as discussed in Ref.~\cite{Skripnikov:2020e}, we demonstrated that by exploiting the properties of the hyperfine interaction operator and the asymptotic behavior of the electronic wave function inside the heavy nucleus, it is possible to combine accurate electronic structure calculations with experimental data to determine the contribution of the magnetization distribution. In addition to the moments derivation, we calculated the electronic EEs for the considered excited electronic states, as well as the IP and electronic $g_J$ factor of the ground electronic state. These parameters can be directly compared with the accurate experimental data. Thus, we can check the reliability of the method and the theoretical uncertainties ascribed.

In the present paper atomic units $\hbar=|e|=m_e=1$ and Gaussian electromagnetic units are used.

\section{Theory}

The hyperfine structure interactions between nuclear moments and electrons can be written in the following way~\cite{johnson2007atomic}:
\begin{equation}
    H_{\rm HFI}=\sum_k T^e_k \cdot T^n_k,
\end{equation}
where $T^e_k$ and $T^n_k$ are irreducible tensors of rank $k$ operating in the space of electronic and nucleus coordinates.

The first-order energy shift of the atomic state $|\gamma IJFM_F\rangle$, characterized by the electronic total angular momentum $J$, nuclear spin $I$, total momentum $F$, its projection on the laboratory axis $M_F$, and remaining quantum numbers $\gamma$, is given by:
\begin{eqnarray}
 W_F^{(1)}=\langle \gamma IJFM_F | H_{\rm HFI} |\gamma IJFM_F\rangle \\ 
          =\sum_k M(IJF,k)  \langle T^e_k \rangle_J \langle T^n_k \rangle_I.
\end{eqnarray}
Here, $\langle T^e_k \rangle_J=\langle J,M_J=J| T^e_{k,0} |J,M_J=J\rangle$ represents the electronic matrix element of the zero spherical component of the $T^e_k$ tensor, where $M_J$ is the projection of $J$ on the laboratory axis. Similarly, $\langle T^n_k \rangle_I=\langle I,M_I=I| T^n_{k,0} |I,M_I=I\rangle$ represents the matrix element for the ``stretched'' nucleus state with the projection $M_I=I$. The coefficients $M(IJF,k)$ are determined by the angular moments algebra, which involves a combination of 6-$j$ and 3-$j$ Wigner symbols.

The commonly-used designations for nucleus moments are as follows: $\langle T^n_1 \rangle_I=\mu_I$ for the nuclear magnetic moment and $\langle T^n_2 \rangle_I=2Q_S$ for the spectroscopic electric quadrupole moment. In terms of these notations, the first-order energy shift induced by hyperfine interaction is parameterized by the following widely-used constants:
\begin{eqnarray}
    \label{A_general}
    A=\frac{\mu}{IJ} \langle T^e_1 \rangle_J, \\
    \label{B_general}
    B=2Q_S \langle T^e_2 \rangle_J=q Q_S. 
\end{eqnarray}
Parameters $A, B$ are the magnetic dipole and electric quadrupole interaction constants, respectively. We also use the notation $q=2\langle T^e_2 \rangle_J$ for the electric field gradient (EFG) at the site of the nucleus.

The explicit expressions for the required electronic operators in the point magnetic dipole approximation are as follows:
\begin{eqnarray}
   \label{T0_point}
   T^e_{1,0} = \sum_i -\frac{i}{r^2} \sqrt{2} \bm{\alpha} \mathbf{C}^{(0)}_{1,0}(\mathbf{r}_i), \\
   T^e_{2,0} = \sum_i -\frac{1}{r^3} C_{2,0}(\mathbf{r}_i).
\end{eqnarray}
Here, $\bm{\alpha}$ is the vector of Dirac matrices, $\mathbf{r}_i$ is the radius vector of the electron $i$ with respect to the position of the nucleus, and $C_{k,0}$ and $\mathbf{C}^{(0)}_{k,0}$ are normalized spherical harmonics and vector spherical harmonics, respectively. Note that the BR effect is automatically accounted for by incorporating in the electronic Hamiltonian the electron-nucleus interaction operator, which includes the finite nuclear charge distribution model. Thus, even when using the point magnetic dipole approximation, the extended charge distribution model is still at work.

To account for the effect of the interaction of electrons with the extended nuclear magnetization, the following modification of Eq.~(\ref{T0_point}) can be used:
\begin{equation}
    \label{hfs1}
    T^e_{1,0} = \sum_i -\frac{i}{r^2} \sqrt{2} \bm{\alpha} \mathbf{C}^{(0)}_{1,0}(\mathbf{r}_i) F(r_i),
\end{equation}
where the
function $F(r)$ encodes the nuclear magnetization distribution within a finite nucleus. In the point magnetic dipole moment approximation, $F(r)$ is equal to 1 for any value of $r$.  The expressions for various models of the extended magnetization distribution can be found in Refs.~\cite{Zherebtsov:2000,Tupitsyn:02,Malkin:2011}.

For an accurate treatment, one should also account for quantum electrodynamics (QED) effects. The magnetic dipole hyperfine interaction constant $A$ can be parameterized as follows:
\begin{equation}
\label{Aparam}
    A = A_0 - A_{\rm BW} + A_{\rm QED}.
\end{equation}
In this equation, $A_0$ represents the HFS constant in the point magnetic dipole approximation, which nevertheless takes into account the finite nuclear charge radius as noted above. $A_{\rm BW}$ corresponds to the contribution of the finite nuclear magnetization distribution effect. $A_{\rm QED}$ denotes the contribution of quantum electrodynamics effects.
For the electronic state $\left| J,M_J=J \right>$ we have:
\begin{eqnarray}
\label{Apar}
 A_0  = \frac{\mu}{I J}\, 
 \left< J,J \right| \sum_i\, \-\frac{i}{r^2} \sqrt{2} \bm{\alpha} \mathbf{C}^{(0)}_{1,0}(\mathbf{r_i}) \left| J,J \right>.
\end{eqnarray}
The finite nuclear magnetization distribution correction $A_{\rm BW}$ is defined as:
\begin{eqnarray}
\label{ABWdefinition}
 A_{\rm BW}  = \frac{\mu}{I J}\, 
 \left< J,J \right| \sum_i\, \-\frac{i}{r^2} \sqrt{2} \bm{\alpha} \mathbf{C}^{(0)}_{1,0}(\mathbf{r_i}) (1-F(r_i)) \left| J,J \right>.
\end{eqnarray}

It was demonstrated in Ref.~\cite{Skripnikov:2020e} that for many-electron heavy atoms and molecules containing such atoms even in states with complex electronic structure,
the contribution of $A_{\rm BW}$ can be factorized as shown in Eq. (29) of Ref.~\cite{Skripnikov:2020e}:
\begin{eqnarray}
 \label{AparBW3}
 A_{\rm BW}
 \approx \frac{\mu}{I J} ({\cal{P}}_s + \beta{\cal{P}}_p)B_s\\
  = \frac{\mu}{I J} \dbtilde{A}_{\rm BWel} B_s, \nonumber
\end{eqnarray}
where we introduced the dimensionless parameter: 
\begin{equation}
\label{ABwexpr}
\dbtilde{A}_{\rm BWel}={\cal{P}}_s + \beta{\cal{P}}_p.
\end{equation}
For a given many-electron wave function, ${\cal{P}}_s$ can be calculated as the difference between the mean values of the projectors on the $1s_{1/2}$ H-like functions with opposite angular momentum projection quantum numbers, truncated at the nuclear charge radius $R_{\rm nuc}$ (see Eqs. (23),(33),(34) of Ref.~\cite{Skripnikov:2020e} for details; we use $R_{\rm nuc}=\sqrt{5/3} r_{rms}$, where $r_{rms}$ is the root-mean-square charge radius of the nucleus in question). ${\cal{P}}_p$ is the similar quantity corresponding to the $2p_{1/2}$ H-like function. The values of the ${\cal{P}}_s$ and ${\cal{P}}_p$ parameters can be considered as indicators of the spin-polarization of the $s_{1/2}$ and $p_{1/2}$ electronic shells inside the nucleus. Therefore, for example, for a system with closed electronic shells, ${\cal{P}}_s$ and ${\cal{P}}_p$ are equal to zero (as well as the whole first-order magnetic dipole HFS constant). Parameter $\beta$ is the electronic factor that is specific  for a given element, but electronic state independent and is determined by the ratio of the products of amplitudes of large and small components of $1s_{1/2}$ and $2p_{1/2}$ functions inside the nucleus. Other functions ($p_{3/2}$, etc.) are not included in the equation because their amplitudes inside the heavy nucleus are very small and according to Eq.~(\ref{ABWdefinition}), the BW effect is quadratic in this amplitude. If we consider a many-electronic state with a single valence electron in the $p_{3/2}$ state, the direct BW effect from this electron will be negligible. However, the effect can appear due to the spin-polarization of $s_{1/2}$ and $p_{1/2}$-type ``closed'' electronic shells~\cite{Schwartz:1955,Persson1998}. This will result in nonzero values of the ${\cal{P}}_s$ and ${\cal{P}}_p$ constants. Parameter $B_s$ is the following matrix element over the H-like $1s_{1/2}$ wave function $\eta_{1s_{1/2,1/2}}$ 
with the total angular momentum $1/2$ and its projection $1/2$:
\begin{equation}
\label{Bs}
B_s=\int\limits_{|\mathbf{r}|\le R_{\rm nuc}} \eta_{1s_{1/2,1/2}}^{\dagger} -\frac{i}{r^2} \sqrt{2} \bm{\alpha} \mathbf{C}^{(0)}_{1,0}(\mathbf{r}) (1-F(r)) \eta_{1s_{1/2,1/2}} d\mathbf{r}. \\
\end{equation} 
The electronic factor $\dbtilde{A}_{\rm BWel}$ defined in Eq.~(\ref{ABwexpr}) is solely determined by the electronic structure and is independent of the model used for the nuclear magnetization distribution. This has been numerically confirmed by considering various models of the nuclear magnetization distribution for atoms and molecules~\cite{Skripnikov:2020e,Prosnyak:2021}. Consequently, all information regarding the nuclear magnetization distribution is included into only one parameter $B_s$. This parameter can be employed to describe through Eq.~(\ref{AparBW3}) the contribution of the magnetization distribution effect for any  electronic state of a many-electron heavy atom, molecule, or compound that contains such an atom.

The considered approach can also be used to approximately calculate the specific difference parameter $\xi$ introduced in a study of Li-like and H-like ions in Ref.~\cite{Shabaev:01a} by using Eq.~(\ref{AparBW3}). The ratio of the parameters $B_s$ estimated for a specific nuclear magnetization distribution model and for a model of a uniformly magnetized ball provides the value of the $d_{\rm nuc}$ parameter introduced in another approach~\cite{konovalova2017calculation,atoms6030039}. Equation (\ref{AparBW3}) ensures that this parameter is independent of the many-electron state of the heavy-atom-containing system, which is typically assumed but has previously only been explicitly explained for single-electron $s_{1/2}$ or $p_{1/2}$ atomic states. Finally, it should be emphasized that in heavy-atom compounds, not only the $s_{1/2}$-type contribution, represented by the ${\cal{P}}_s$ term in Eq.~(\ref{AparBW3}) is important to describe the BW effect as may be the case for some systems, but also the $p_{1/2}$ contribution, represented by the $\beta{\cal{P}}_p$ term in Eq.~(\ref{AparBW3}) should be considered. For example, neglecting the latter term would lead to the loss of the BW effect in the $7p_{1/2}$ state of Ra$^+$~\cite{Skripnikov:2020e} in contradiction with a direct calculation within a certain model.

The parameter $B_s$ has a well-defined physical meaning~\cite{Skripnikov:2020e,Skripnikov:2022}. For the $1s$ state of a H-like ion, $\dbtilde{A}_{\rm BWel}=1$ as follows from Eq.~(\ref{ABwexpr}). Consequently, we have
\begin{equation}
\label{BsHlike}
B_s= \frac{I}{2 \mu}  A_{\rm BW}({\rm 1s, H\text{-}like~ion}).
\end{equation}
This means that $B_s$ determines the contribution of the finite nuclear magnetization distribution effect to the magnetic HFS constant $A$ of the H-like ion. If the nuclear magnetic moment was measured by the method which is not connected with the HFS analysis then using electronic structure parameters that can be computed with high precision for H-like ions~\cite{shabaev1997ground,Shabaev:01,Volotka:2008}, it becomes feasible to extract $B_s$ with high accuracy~\cite{Skripnikov:2022}.

By introducing $\Tilde{A}_0=A_0 / \frac{\mu}{I J}$, $\Tilde{A}_{\rm QED}=A_{\rm QED} / \frac{\mu}{I J}$ one can rewrite Eq.~(\ref{Aparam}) as follows:
\begin{equation}
\label{Aparam2}
    A = \frac{\mu}{I J}(\Tilde{A}_0 - \dbtilde{A}_{\rm BWel} B_s + \Tilde{A}_{\rm QED}).
\end{equation}
Parameters $\Tilde{A}_0, \dbtilde{A}_{\rm BWel}$ and $\Tilde{A}_{\rm QED}$ do not depend on the magnetic properties of the nucleus. In the present work, we use parametrization of Eq.~(\ref{Aparam2}) as it allows us to separate these purely electronic parameters and the nuclear parameter $B_s$ without referencing any nuclear magnetization model.

To calculate $B_s$ one has to know function $F(r)$. Many nuclear factor calculations ($B_s$ in our case) are based on the single-particle approximation without the uncertainty control. However, the use of the parametrization (\ref{Aparam2}) allows one to avoid direct calculation of $B_s$ by combining calculated electronic factors with the experimental data. Such an approach was used to extract $B_s$ using the experimental HFS $A$ constant for the ground state of Ra$^+$:
\begin{equation}
\label{BsExpression}
    B_s=\frac{A_0 +A_{\rm QED} - A}{\frac{\mu}{I J} \dbtilde{A}_{\rm BWel}},
\end{equation}
and the nuclear magnetic moment of the $^{225}$Ra measured independently of the HFS studies. Good agreement of the predicted value of the excited state HFS constant (where the extracted $B_s$ values was used) with experiment justifies this approach. When the HFS $A$ constant was measured for at least two electronic states with different configurations, one can calculate $B_s$ factor without rarely accessible information on the independently measured nuclear magnetic moments. Namely, according to Eq.~(\ref{Aparam2}), the ratio of these $A$ constants is independent of the nuclear magnetic moment and the value of $B_s$ can be extracted from this ratio. With the obtained $B_s$ value the nuclear magnetic moment can be determined by the formula:
\begin{equation}
    \label{muFormula}
    \mu_I = \frac{A I J}{\Tilde{A}_0 + \Tilde{A}_{\rm QED} - \dbtilde{A}_{\rm BWel} B_s}.
\end{equation}
The same approach was used in Ref.~\cite{Porsev:2021} to extract $\mu$($^{229}$Th). The similar procedure for the determination of the model-independent HFS anomaly with the use of the ratio of two HFS constants was implemented in Refs.~\cite{Ehlers:1968,barzakh2012hyperfine,Schmidt:2018} (see also Ref.~\cite{Persson1998}).

To conclude the discussion on the separation of pure electronic and nuclear magnetization distribution effects, one can establish the following relationship between the atomic screening factors $x^s_{scr}$ and $x^p_{scr}$ introduced in Ref.~\cite{Ginges:2022} for the case of atoms, specifically in the $s_{1/2}$ and $p_{1/2}$ states, with our parametrization~\cite{Skripnikov:2020e} outlined above: $x^s_{scr} = -{\cal{P}}_s \Tilde{A}_0(1s,{\rm H-like}) / \Tilde{A}_0$ and $x^p_{scr} = -{\cal{P}}_p \Tilde{A}_0(2p_{1/2},{\rm H-like}) / \Tilde{A}_0$.

For the Po atom, the electronic $g_J$ factor in the ground electronic state was measured~\cite{axensten1961nuclear}. 
For an atom with a spinless nucleus, the first-order Zeeman shift for the electronic state with the total angular momentum $J$ and its projection $M_J$ is directly related to the electronic $g_J$ factor:
\begin{equation}
 \label{Zeeman1}
   \Delta E^{(1)} = g_J M_J \mu_0 B,
\end{equation}
where $B$ is the value of the external magnetic field and $\mu_0$ is the Bohr magneton. Thus, the theoretical value of the electronic $g_J$ factor can be deduced from the derivative of the calculated electronic energy with respect to the applied external magnetic field at the point of zero field. In the four-component Dirac theory, this interaction is described by the following Hamiltonian:
\begin{equation}
 \label{Zeeman2}
   H_Z=\mu_0\sum_i[\bm{r_i}\times\bm{\alpha_i}] \cdot \bm{B},
\end{equation}
where $i$ is an electron index and summation goes over all electrons in the system. The leading contribution of QED effects to the atomic magnetic moment (and $g_J$ factor) outside the Breit approximation can be approximately estimated as an expectation value of the following operator~\cite{Cheng:85}:
\begin{equation}
 \label{QEDeq}
   \mu_0\frac{g_e-2}{2}\sum_i\beta_i\Sigma_{z,i},
\end{equation}
where
$\beta=
\left(\begin{array}{cc}
1 & 0 \\
0 & -1 \\
\end{array}\right), 
$ 
$\Sigma_{z}$ is the $z$ component of the vector operator
$\bm{\Sigma}=
\left(\begin{array}{cc}
\bm{\sigma} & 0 \\
0 & \bm{\sigma} \\
\end{array}\right)\ ,
$
$\bm{\sigma}$ are the Pauli matrices, and $g_e=2.0023193\dots$ is the free-electron $g$ factor.

\section{Calculations}

In this section, we will describe the scheme and technical details of the electronic structure calculations. In a dedicated subsection, we will outline the systematic scheme that was employed to estimate the uncertainty of these calculations.

\subsection{Computational scheme}

To account for the leading relativistic and electronic correlation effects in a balanced way, we performed calculations using the relativistic coupled cluster method (CC) with single, double, and perturbative triple excitation amplitudes CCSD(T)~\cite{Visscher:96a,Bartlett:2007}, employing the Dirac-Coulomb Hamiltonian. In this calculation, all electrons of the Po atom were included in the correlation treatment, and the energy cutoff for virtual orbitals was set to 10,000 hartree. The choice of such a large cutoff is crucial to properly consider the correlation contributions of the inner-core electrons~\cite{Skripnikov:17a,Skripnikov:15b}. We used the Dyall's uncontracted augmented all-electron quadruple-zeta AAE4Z basis set~\cite{Dyall:06,Dyall:12}, which was extended by additional diffuse and tight Gaussian functions. In total, the basis set consisted of 39$s$-, 35$p$-, 28$d$-, 22$f$-, 10$g$-, 5$h$-, and 4$i$-type Gaussian functions. This can be succinctly expressed as [39,35,28,22,10,5,4]. The HFS constants and $g_J$ factor were calculated using the finite-field approach.

To account for higher-order correlation effects, we incorporated the following corrections:

(i) The ``CCSDT-3 $-$ CCSD(T)'' correction. This correction involved comparing the values of corresponding parameters computed within the CCSDT-3~\cite{Noga:CCSDT-3:87} and the CCSD(T) models. In the CCSDT-3 model, triple excitation amplitudes are treated iteratively, while in the CCSD(T) approach, they are considered noniteratively. However, the CCSDT-3 model still lacks some diagrams presented in the full CCSDT model~\cite{Bartlett:2007}. We calculated this correction for 56 correlated electrons of Po using the [35,29,21,14,6] basis set, which is the extension of the AE3Z basis set~\cite{Dyall:06,Dyall:12} and includes additional diffuse and tight functions of $s$-, $p$-, $d$-, and $f$-types, as well as reoptimized $g$-type basis functions within the procedure developed in~\cite{Skripnikov:2020e,Skripnikov:13a}. 
Due to the reduced number of the inner-core electrons included in correlation treatment in this and all other 56-electron correlation calculations the energy cutoff for virtual orbitals was set to 300 hartree.
For the purpose of the uncertainty estimation, we additionally computed the correction using the [35,29,21,14,4,1] basis set within the 56-electron CCSD(T) approach and the original set of $g$-type functions of the AE3Z basis. Furthermore, we evaluated the ``CCSDT-3 $-$ CCSD(T)'' correction using the expanded [35,29,21,14,6,3] basis set but with a reduced number of correlating electrons -- 48 instead of 56.

(ii) The ``CCSDT $-$ CCSDT-3'' correction was determined as the difference between results obtained within the full CCSDT and CCSDT-3 models. This correction was computed for 24 valence and outer-core electrons of Po using the [35,29,21,14,4,1] basis set. In this and other 24-electron correlation calculations the energy cutoff for virtual orbitals was set to 20 hartree.

(iii) The ``CCSDT(Q) $-$ CCSDT'' correction was calculated as the difference between results obtained within the CCSDT(Q)~\cite{Kallay:6} and CCSDT models employing the same basis set and the same number of correlated electrons as in the previous correction. The CCSDT(Q) model incorporates coupled cluster with single, double, triple, and perturbative quadruple excitation amplitudes, thereby introducing the contribution of connected quadruple excitation amplitudes. 

To account for the effects of a larger basis set, we computed the following corrections:

(i) Correction for an increased number of $f$-type functions: We added 6 $f$-type functions and calculated this correction for the 56 outer electrons.

(ii) Correction for an extended number of high-angular momentum basis functions: This correction was determined by comparing the results obtained using the CCSD(T) method with two different basis sets. The first calculation employed the [39,35,28,22,12,9,8] basis set, while the second calculation used the [39,35,28,22,10,5,4] basis set. In this case, 56 electrons were correlated.

(iii) We also computed the extrapolated contribution of basis functions with $L>6$, similar to the approach described in Refs.~\cite{Skripnikov:2021a,AthanasakisRaFPinning:2023}. This calculation was performed using the CCSD(T) method and the Dirac-Coulomb Hamiltonian.

We also considered the correction due to the higher-level electronic structure Hamiltonian. We calculated the contribution of the Gaunt interelectron interaction using the molecular-mean-field exact two-component approach~\cite{Sikkema:2009} at the Fock-space coupled cluster method with single and double excitation amplitudes (FS-CCSD). Finally, we estimated the contribution of the QED effects. Vacuum-polarization effects were taken into account by including the Uehling potential, and the self-energy contribution was estimated using a model self-energy Hamiltonian approach~\cite{Shabaev:13,Malyshev:2022} reformulated for molecular and atomic calculations~\cite{Skripnikov:2021a}. 
The sum of these QED contributions will be referred to as the model QED contribution below.
Correction on these QED effects were calculated within the all-electron CCSD(T) method without energy cutoff. To account for QED effects in the calculation of the electronic $g_J$ factor, we additionally used the operator of Eq.~(\ref{QEDeq}), which has been shown to work well in the similar CC calculations~\cite{Maison:2019}.

In addition to the aforementioned corrections, we also calculated the contribution of virtual orbitals with energies above 10,000 hartree, referred to as the ``High virt.'' correction for HFS constants. For this correction, we employed the [35,29,21,14,6,3,2] basis set and the 84-electron relativistic CCSD(T) method. 

All calculations described above were performed for the root-mean-square charge radius equal to $5.53$~fm according to the interpolation formula from Ref.~\cite{Johnson:1985}. We used the Gaussian nuclear charge distribution model~\cite{Visscher:1997} in the majority of calculations. However, a more realistic Fermi nuclear charge distribution model was also considered for the HFS constants. The difference between the values calculated using the Fermi and Gaussian nuclear charge distribution models was determined using the 84-electron CCSD(T) approach without a virtual energy cutoff.

Relativistic electronic structure calculations were performed using the {\sc dirac}~\cite{DIRAC19,Saue:2020} and {\sc mrcc}~\cite{MRCC2020,Kallay:1,Kallay:2} codes. The FS-CCSD calculations were performed using the {\sc exp-t}~\cite{EXPT_website,Oleynichenko_EXPT,Zaitsevskii:2023} code. Scalar relativistic correlation calculations used for test purposes and for the basis set construction were carried out using the {\sc cfour}~\cite{CFOUR,Matthews:CFOUR:20} code. To generate a compact set of basis functions with high angular momenta, we used the code developed in Refs.~\cite{Skripnikov:2020e,Skripnikov:13a}. The code for calculating matrix elements of considered operators over atomic bispinors was developed in Refs.~\cite{Skripnikov:16b,Schmidt:2018}. We also used the code developed in Ref.~\cite{Skripnikov:2021a} to incorporate self-energy effects within the model QED approach~\cite{Shabaev:13,Malyshev:2022} and vacuum-polarization effects. To calculate the correction on the nuclear charge distribution model, we developed a code that computed integrals of the Fermi nuclear charge distribution potential over the Gaussian-type basis functions employed in the present study.

\subsection{Uncertainty estimation}

The calculation scheme described in the previous subsection includes the reference (main) calculation, followed by a set of corrections aimed at improving the consideration of correlation effects, extending the basis set, and refining the Hamiltonian. The reference calculation was chosen to provide a balanced and accurate initial description of the many-electron system under consideration. Thus, the magnitudes of the corrections are expected to be small (see the next section). At the same time, in the framework of the chosen reference approach we were working close to the limits of the available resources. Below, we estimate theoretical uncertainties and conservatively assume that all higher-order contributions is comparable to the contribution at the last implemented step when we have no other bases for uncertainty estimation. Although each of our estimations should be considered as tentative, one can confirm the overall adequacy of the ascribed total uncertainties by the comparison with the experimental data for the EEs, IP, and $g_J$ factor.

We considered the following sources of uncertainties:

(i) As mentioned earlier, the correction ``CCSDT-3 $-$ CCSD(T)'' was calculated in three different ways. As the uncertainty estimation we took the difference between the largest and the smallest of these values. Note, that the typical uncertainty for this correction is about 30\% (see tables in the next section).

(ii) Due to its complexity, the ``CCSDT $-$ CCSDT-3'' correction was calculated within one way. 
However, similar to the previous case of the ``CCSDT-3 - CCSD(T)'' correction, the aim of the ``CCSDT - CCSDT-3'' is to improve the treatment of triple excitation amplitudes. Therefore, it can be assumed that these two corrections would have similar uncertainties (i.e. about 30\%). However, to be conservative, we arbitrarily increased the uncertainty to 50\%.

(iii) The total value of the ``CCSDT(Q) $-$ CCSDT'' correction, which accounts for the effect of quadruple excitation amplitudes, can be considered as an estimate of the unaccounted correlation effects of the higher order. Therefore, we set the uncertainty of this correction as 100\% of its value.

\begin{table*}
\caption{Calculated values of the ionization potential IP and excitation energies EE of the $6p^4~{}^1D_2$ and $6p^3 7s^1~{}^5S_2$ electronic states. In the last column, the value of the electronic $g_J$ factor for the ground electronic state $6p^4~{}^3P_2$ is given.}

\begin{tabular*}{1.0\textwidth}{l@{\extracolsep{\fill}}llll}
\hline
\hline 
Method              & IP            & EE($6p^4~{}^1D_2$) (cm$^{-1}$)     & EE($6p^3 7s^1~{}^5S_2$) (cm$^{-1}$)     & $g_J$($6p^4~{}^3P_2$)   \\
\hline                                                
CCSD(T)             & 67753         & 21873           & 38956         & 1.39336            \\
CCSDT-3 $-$ CCSD(T) & -56(24)       & 63(3)           & -76(27)       & 0.00253(49)     \\
CCSDT $-$ CCSDT-3   & -109(54)      & -10(5)          & -108(54)      & $< 1\cdot10^{-5}$           \\
CCSDT(Q) $-$ CCSDT  & -21(21)       & -32(32)         & -46(46)       & -0.00010(10)       \\
Basis set correction& 314(86)       & -5(6)           & 303(64)       & -0.00044(29)       \\
Gaunt               & -27(14)       & -211(105)       & -33(17)       & 0.00015(15)        \\
QED                 & 23(7)         & 17(5)           & 58(17)        & 0.00087(4)        \\
Total               & 67878(107)    & 21696(111)      & 39053(102)    & 1.39637(59)      \\
Experiment          & 67896.310(44)~\cite{Po_IPexp:2019} & 21679.11(1)~\cite{Charles:66} & 39081.19(1)~\cite{Charles:66}     & 1.39609(4)~\cite{axensten1961nuclear}          \\
\\
                    & \multicolumn{4}{l}{Other theory:} \\
SCF~\cite{Peterson:2003}            & 69156           &       &   &   \\                  
DK+CASPT2+SO~\cite{Roos:2004}       & 66863           & 20003 &   &   \\
4c-MBPT~\cite{Zeng:2010}            & 66476           & 20282 &   &    \\
rp-ccCA~\cite{Wilson:2012}          & 68342           &       &   &    \\
CCSD(T)+Breit+QED~\cite{Borschevsky:2015}~~~ & 68008  &       &   &    \\
\hline
\hline
\end{tabular*}
\label{TValProps}
\end{table*}

(iv) As described above, the correction on the extension of the basis set includes three components. Two of them correspond to the direct calculation of the effect of an increased basis set by some set of functions, and the remaining one corresponds to the extrapolated contribution. The uncertainty of each of the two corrections was estimated by comparing their contributions calculated within the CCSD(T) and CCSD models. The extrapolated scheme has demonstrated good performance for energies~\cite{Skripnikov:2021a} using the same electron correlation methods as employed here. However, to be conservative, the uncertainties of the extrapolated contributions to IP and EE were calculated as the difference between the results obtained within the CCSD(T) and CCSD models, but with an additional doubling. To be more conservative in the cases of HFS constants and $g_J$ factor, we set the uncertainty of the extrapolated contribution to 100\% of its value.

(v) In the present work, we only considered the contribution of the Gaunt term of the full Breit interelectron interaction. The Gaunt term is usually the dominant one ($\approx$~80~\%) for energies~\cite{Skripnikov:2021a}. Therefore, for EEs and IP, we suggest that the contribution of the remaining part of the Breit interaction can be 50\% of the Gaunt term. However, to be more conservative, and taking into account possible negative-energy states contribution~\cite{Maison:2019} which was neglected, we consider a 100\% uncertainty for the Gaunt interaction term in the case of HFS and $g_J$ factor calculations.

(vi) To account for the contribution of QED effects for IP, EEs and HFS constants, we used the model QED approach~\cite{Shabaev:13,Malyshev:2022,Skripnikov:2021a}. This approach is quite accurate for electronic energy calculations, but it is not complete for treating hyperfine structure constants~\cite{Shabaev:13}. Tests of the model QED approach on different atomic ions~\cite{Shabaev:13,Malyshev:2022,Skripnikov:2021b} verify its high accuracy for the energies calculations, achieving just a few percent deviation from the rigorous QED treatment. Even in systems with a significant cancellation of QED effects, where the tiny interference between QED and electron correlation effects is important, the maximum deviation from the rigorous QED treatment was found to be 14\%~\cite{ShabaevMolQED:2020}. To be even more conservative, we set the uncertainty of this correction as 30\% for IP and EEs. For the HFS constants, we set the uncertainty of this contribution to 100\% due to the mentioned above incomplete consideration of QED terms in the model QED approach in the HFS case. For the case of the electronic $g_J$ factor, we used the operator presented in Eq.~(\ref{QEDeq}) to account for the QED effects, which refer to the free-electron $g$ factor. In case of B-like Ar ion with the valence $p_{1/2}$ or $p_{3/2}$ electronic configuration we found~\cite{Maison:2019} that the deviation of such QED treatment from the rigorous QED approach is less than 1\%. In the present work, we found that the QED correction to the $g_J$ factor of Po, calculated at the CCSD and CCSD(T) levels, deviates by just 0.5\% of the value of this correction. We also calculate an additional QED contribution to $g_J$ factor by including the model QED operator into the electronic Hamiltonian (as in case of HFS). To be conservative (in the absence of the rigorous QED results), we suggest that the overall uncertainty of the QED contribution to $g_J$ factor is 5\%.

(vii) In calculations, we used the Gaussian and Fermi models of the nuclear charge distribution, which provide quite distinct descriptions of the nuclear charge density, much more pronounced than the difference between the Fermi and uniformly charged ball models, which is often considered as a measure of the uncertainty arising from the nuclear charge distribution~\cite{maartensson2003atomic}. We propose that the uncertainty from the choice of nuclear charge model is smaller than half of the discrepancy between the results obtained using the Gaussian and Fermi models.
As a nuclear radius we used the root-mean-square charge radius of $^{202}$Po ($5.53$ fm) calculated by empirical formula fitted to the measured radii~\cite{Johnson:1985}. In principle, there is a slight dependence of the $\Tilde{A}_0$ parameter on the radius. To investigate this effect, we increased the nuclear radius by about 0.13~fm, which corresponds to a difference of $\delta \langle r^2 \rangle^{218,192} \approx 1.5$ fm$^2$~\cite{Cocolios:2011}. The obtained correction is more than an order of magnitude smaller than the uncertainty of the $\Tilde{A}_0$ constant, though it is accounted for in our uncertainty treatment. 

(viii) Finally, we set the uncertainty of the contribution of high-lying virtual orbitals (with energies greater than 10,000 hartree) to be 100\%.

The total uncertainty was estimated as the square root of the sum of squares of all uncertainties described above.

\section{Results and discussion}

The calculated values of the IP and EEs of the electronic states of interest ($6p^4~{}^1D_2$ and $6p^3 7s^1~{}^5S_2$) as well as the value of the electronic $g_J$ factor in the ground electronic state are given in Table~\ref{TValProps}.

These values perfectly agree with experiment~\cite{Po_IPexp:2019,Charles:66}. Their deviations from the measured values are: 18 cm$^{-1}$ (0.03\%) for IP, 17 cm$^{-1}$ (0.08\%) for the EE of the $6p^4~{}^1D_2$ state and 28 cm$^{-1}$ (0.07\%) for the EE of the $6p^3 7s^1~{}^5S_2$ state. As one can see from Table~\ref{TValProps} the agreement with experiment is better than in all previous theoretical studies. 

The calculated value of the electronic $g_J$ factor for the ground electronic state deviate from the experiment~\cite{axensten1961nuclear} by 0.02\%. Interestingly, as it follows from Table~\ref{TValProps}, QED effects give a quite significant contribution compared to other corrections. The neglect of QED effects would worsen the agreement with the experiment. The good agreement obtained for the electronic $g_J$ factors allows us to probe the accuracy of accounting for spin-orbit interactions for the electronic wave function and correlation effects. 
An accurate treatment of both these effects in the electronic wave function is important for accurately describing the HFS constants~\cite{olsmats1961hyperfine}. In the absence of the spin-orbit interaction, the electronic $g_J$ factor is determined by the standard Lande formula~\cite{LL77} giving a value of $g_J=1.5$, which is significantly deviate from the experimental value $g_J=1.39609(4)$~\cite{axensten1961nuclear}.

Note that our theoretical uncertainty proves to be quite conservative for the IP, EEs, and  electronic $g_J$ factor. This gives confidence in our theoretical uncertainties estimation for the calculated HFS  constants.

The calculated values of the constants $\Tilde{A}_0$ and $\Tilde{A}_{\rm QED}$, EFG for the ground $6p^4~{}^3P_2$ and excited $6p^4~{}^1D_2$ and $6p^3 7s^1~{}^5S_2$ electronic states of Po are presented in Table~\ref{TResults}.

The hyperfine structure constants $A$ and $B$ for the ground $6p^4~{}^3P_2$ and excited $6p^3 7s^1~{}^5S_2$ electronic states of $^{205}$Po and $^{207}$Po are known with high accuracy (see Refs.~\cite{olsmats1961hyperfine,Kowalewska:1991}). In the following analysis, we will use these data to determine nuclear moments for $^{205}$Po and $^{207}$Po and deduce the BW effect for them.

\subsection{Magnetic dipole moments and BW effect in Po isotopes}

\begin{table*}
\caption{The calculated values of the magnetic dipole HFS constants $\Tilde{A}_0$, $\Tilde{A}_{\rm QED}$, EFG $q$ for the ground and excited electronic states. }
\begin{tabular*}{1.0\textwidth}{l@{\extracolsep{\fill}}lllccc}
\hline
\hline             
                    & \multicolumn{3}{c}{ $\Tilde{A}_0$ (MHz)}                                      & \multicolumn{3}{c}{$q$ (a.u.)}     \\
Method              & $6p^4~{}^3P_2$   & $6p^4~{}^1D_2$   & $6p^3 7s^1~{}^5S_2$  & $6p^4~{}^3P_2$ &    $6p^4~{}^1D_2$ & $6p^3 7s^1~{}^5S_2$         \\
\hline                                                                                                                                        
CCSD(T)             & 806              & 8806             & 3683                 & 5.479               & -16.74              & 5.858            \\
CCSDT-3 $-$ CCSD(T) & -22(6)           & 3(5)             & -119(25)             & 0.290(54)           &  0.10(4)            & -0.386(75)       \\
CCSDT $-$ CCSDT-3   & 6(3)             & -4(2)            & 43(21)               & -0.015(8)           &  0.01(1)            & -0.012(6)        \\
CCSDT(Q) $-$ CCSDT  & 4(4)             & -2(2)            & 18(18)               & -0.010(10)          &  0.03(3)            & -0.033(33)       \\
Basis set correction& 21(12)           & 19(32)           & 34(14)               & -0.031(9)           & 0.05(1)             & 0.066(57)        \\
Gaunt               & -14(14)          & -84(84)          & -15(15)              & -0.037(37)          & 0.12(12)            & -0.053(53)       \\
Nucl. charge model  & 2(3)             & -2(3)            & -3(5)                &  0.000(4)           & $< 1\cdot10^{-2}$   &  0.002(1)        \\
High. virt.         & -5(5)            & 1(1)             & 2(2)                 & -0.004(4)           & 0.01(1)             & -0.007(7)        \\
QED                 & 16(16)$^a$       & 9(9)$^a$         & -28(28)$^a$          & -0.005(5)           & 0.02(2)             &  $< 1\cdot10^{-3}$  \\
Total               & 814(26)$^b$      & 8745(91)$^b$     & 3614(51)$^b$         & 5.666(68)           & -16.39(13)          & 5.433(114)       \\
\hline
\hline
\end{tabular*}
\begin{flushleft}
$^a$ This is the value of $\Tilde{A}_{\rm QED}$.\\
$^b$ This is the value of $\Tilde{A}_0+\Tilde{A}_{\rm QED}$.
\end{flushleft}
\label{TResults}
\end{table*}

The ratio of the HFS constants $A$ for the ground and excited electronic states for a given isotope is independent of the nuclear magnetic moment value, though this ratio can depend on the nuclear magnetization distribution (see Sec. II). Using $A$-constant values from~\cite{olsmats1961hyperfine,Kowalewska:1991} we obtain the corresponding ratios for the $^{205}$Po and $^{207}$Po isotopes:
\begin{eqnarray}
    \label{data205}
    \frac{A(6p^4~^3P_2)}{A(6p^3 7s^1~^5 S_2)} (^{205}{\rm Po})=0.2484(12), \\
    \label{data207}
    \frac{A(6p^4~^3P_2)}{A(6p^3 7s^1~^5 S_2)} (^{207}{\rm Po})=0.2474(4).
\end{eqnarray}
Following Eq.~(\ref{Aparam2}) and combining these ratios with the calculated values of $\Tilde{A}_0+\Tilde{A}_{\rm QED}$ given in Table \ref{TResults} and calculated values of $\dbtilde{A}_{\rm BWel}$ parameters, 
\begin{eqnarray}
  \label{Abwel_gs}
  \dbtilde{A}_{\rm BWel}(6p^4~^3P_2)=-8.98(18) \times 10^{-6}, \\
  \label{Abwel_s}
  \dbtilde{A}_{\rm BWel}(6p^3 7s^1~^5 S_2)=1.90(4) \times 10^{-5}, 
\end{eqnarray}
we can extract the value of the $B_s$ constants for $^{205}$Po and $^{207}$Po isotopes:
\begin{eqnarray}
   \label{Bs205}
   B_s(\rm ^{205}{\rm Po})=6.1(2.1)\times 10^6~{\rm MHz}, \\
   \label{Bs207}
   B_s(\rm ^{207}{\rm Po})=5.8(2.1)\times 10^6~{\rm MHz} .   
\end{eqnarray}
For both electronic states, $\dbtilde{A}_{\rm BWel}$ is strongly dominated by the ${\cal{P}}_s$ term in Eq.~(\ref{ABwexpr}).
In terms of the widely used BW parameter $\epsilon$,
\begin{equation}
\label{BWeps}
\epsilon=A_{\rm BW}/A_0\approx \frac{\dbtilde{A}_{\rm BWel}}{\Tilde{A}_0} B_s,   
\end{equation}
we have: $\epsilon(6p^4~{}^3P_2,^{205}{\rm Po})=-6.7(2.4)\%$, $\epsilon(6p^3 7s^1~^5 S_2,^{205}{\rm Po})=+3.2(1.1)\%$. These results are quite reasonable. In the $6p^3 7s^1~^5 S_2$ state, the BW effect is mainly determined by the valence $7s_{1/2}$ electron, which has a non-negligible amplitude inside the nucleus. Therefore, we obtain the usual sign for $\epsilon$, which corresponds to a reduction in the magnetic dipole HFS constant. For the ground $6p^4~^3P_2$ electronic state, the BW effect is negligible in the zero-order Dirac-Hartree-Fock approximation. This is because both valence open-shell electrons are of the $6p_{3/2}$ nature and have negligible amplitude inside the nucleus. However, due to correlation effects and spin-polarization of the core electrons, we obtain a large value for $\epsilon$, which has the opposite sign compared to the states with a valence $s$ electron. According to our calculations, strong correlation effects reduce the value of the $A_0$ constant for the ground electronic state by a factor of 3.1. A similar effect was also observed and analyzed at the CC level for the $6P_{3/2}$ state of the neutral Tl atom~\cite{Prosnyak:2020,Prosnyak:2021}.

In terms of the $d_{\rm nuc}=B_s/B_s^{(Ball)}$ parameter we obtain: $d_{\rm nuc}(^{205}{\rm Po})=1.74(61)$ and $d_{\rm nuc}(^{207}{\rm Po})=1.67(60)$. This result can be compared with the value calculated within the simple single-particle model of the nuclear magnetization distribution~\cite{bohr1950influence,shabaev1994hyperfine} for the nuclei with a single valence neutron in the $f_{5/2}$ shell-model state: $d_{\rm nuc}(\nu f_{5/2})=1.3$. The simply estimated values are equal for both isotopes, as in both cases the valence nucleon is the neutron in the $f_{5/2}$ state~\cite{olsmats1961hyperfine}. Thus, the directly extracted value and the result of the oversimplified single-particle estimation do not contradict each other in the limits of the ascribed uncertainties.

Taking into account the calculated factors as well as the experimental values of $A$ constants for the $6p^3 7s^1~{}^5S_2$ state~\cite{Kowalewska:1991}, one obtains the magnetic moments (see Eq.~(\ref{muFormula})):
\begin{eqnarray}
    \label{mu205}
    \mu(\rm ^{205}{\rm Po})=0.775(15)~\mu_\mathit{N}, \\
    \label{mu207}
    \mu(\rm ^{207}{\rm Po})=0.805(15)~\mu_\mathit{N}.
\end{eqnarray}
These values are in good agreement with the values measured in the NMR experiment~\cite{herzog1983nuclear}: $\mu(\rm ^{205}{\rm Po})=0.760(55)~\mu_\mathit{N}$ and $\mu(\rm ^{207}{\rm Po})=0.793(55)~\mu_\mathit{N}$, but with the four times smaller uncertainty. Note, however, that they are not consistent with the tentative values deduced in Ref.~\cite{olsmats1961hyperfine} using a semiempirical calculation of the electronic factor of the magnetic dipole interaction $\Tilde{A}_0$: $\mu(\rm ^{205}{\rm Po})=0.26~\mu_\mathit{N}, \mu(\rm ^{207}{\rm Po})=0.27~\mu_\mathit{N}$. These values are underestimated by a factor of 3 compared to our values. It was stated in Ref.~\cite{olsmats1961hyperfine} that the magnetic moment values are very uncertain, with an estimated uncertainty of about $1~\mu_\mathit{N}$. In that work, the electronic factor $\Tilde{A}_0$ was estimated under the approximation that only $6p$ electrons contribute to the electronic factor. However, according to our estimates, the effects of correlation of the ``core'' electrons of Po ($1s^22s^22p^63s^23p^63d^{10}4s^24p^64d^{10}4f^{14} 5s^25p^65d^{10}6s^2$) in the ground electronic state reduce the value of the $A_0$ constant (and correspondingly increase the value of the deduced magnetic moment) by a factor of 2.4. Furthermore, as mentioned earlier in relation to the Dirac-Hartree-Fock level, the contribution of correlation effects is even larger, resulting in a factor of 3.1. Therefore, it appears that this is the main effect that was not considered in Ref.~\cite{olsmats1961hyperfine} and this is the reason of the large difference between our results and that of Ref.~\cite{olsmats1961hyperfine}.

The experimental data for the HFS $A$ constants are available for $^{209}$Po~\cite{Kowalewska:1991}, as well as for $^{193ls, 193hs, 195ls, 195hs, 197, 197m}$Po, $^{199, 199m, 201, 201m, 203, 203m, 211}$Po~\cite{Seliverstov:2014_Po}, and $^{217}$Po~\cite{Fink:2015}, but only for the $6p^3 7s^1~^5 S_2$ state. This means that we cannot extract the values of the $B_s$ constant as we did for $^{205,207}$Po. 
However, we can get an idea of the $B_s$ behavior from the measured HFS anomaly in isotonic Hg nuclei. Indeed, spin of the Hg isotopes and isomers with the neutron number in the range between 113 and 123 takes the same values as in the corresponding Po nuclei ($1/2$, $3/2$, $5/2$, $13/2$)  and the nuclear structures of the isotones with the same spin are nearly identical. Therefore, one can expect that the $B_s$ values for the similar Po and Hg nuclei will be close to each other. Taking into account calculations for Hg nuclei in Ref.~\cite{Reimann:1973} one obtains that $B_s(I)/B_s(5/2)$ gets the values between 0.7 and 1.3 ($I=1/2, 3/2, 13/2$). Single-particle estimation gives the values in the same interval. Thus, one can use the conservative estimation $B_s(I)/B_s(5/2)=1.0(4)$ in the magnetic moment evaluation by Eq.~(\ref{muFormula}). The values of the magnetic moments derived from the measured HFS $A$ constants~\cite{Seliverstov:2014_Po,Fink:2015,Wouters:1991} are given in Table~\ref{TResultsMoments}. Now, for the majority of the short-lived isotopes under consideration, the uncertainty in the deduced magnetic dipole moments is dominated by the experimental uncertainties. Due to the reevaluation of electronic factors, the total uncertainty given in  Refs.~\cite{Seliverstov:2014_Po,Fink:2015} for magnetic moments has been significantly reduced.

Finally, we would like to note that our calculation scheme is well-suited for accurately describing the $6p^4~{}^1D_2$ excited electronic state.  According to Table \ref{TResults}, the theoretical uncertainty of the calculated electronic factors is smaller for this state compared to other considered states. The calculated value of the $\dbtilde{A}_{\rm BWel}$ parameter is:
\begin{equation}
    \dbtilde{A}_{\rm BWel}(^1D_2)=1.26(3) \times 10^{-5}.
\end{equation}
This state has the smallest HFS anomaly: In terms of the $\epsilon$ parameter (see Eq.~(\ref{BWeps})) we have: $\epsilon(^1D_2)/\epsilon(^3P_2)=-0.13$, $\epsilon(^1D_2)/\epsilon(^5 S_2)=-0.27$. Consequently, the effect of the differential hyperfine magnetic anomaly will be reduced for this state. In contrast to the ground state and the $6p^3 7s^1~{}^5S_2$ excited state, $\dbtilde{A}_{\rm BWel}(^1D_2)$ is dominated by the $\beta{\cal{P}}_p$ term in Eq.~(\ref{ABwexpr}). The contribution from the ${\cal{P}}_s$ term is approximately three times smaller and has an opposite sign. Therefore, the $6p^4~{}^1D_2$ state is an interesting example, where both BW contributions are important and partially cancel each other.
Apart from the marked decrease of the poorly known BW contribution, relative theoretical uncertainty of the deduced magnetic moment stemming from the uncertainty of the $\Tilde{A}_0+\Tilde{A}_{\rm QED}$ calculation is noticeably lower (1.4 times) for the $6p^4~{}^1D_2$ state in comparison with the $6p^3 7s^1~{}^5S_2$ state.
Potentially, the accuracy of nuclear magnetic moments could be further improved by conducting new measurements of the hyperfine structure of the $6p^4~{}^1D_2$ state and combining it with the calculated electronic factors. Note, however, that as mentioned earlier, at present, the uncertainty in deducing magnetic moments for the majority of Po isotopes is primarily dominated by experimental uncertainties rather than uncertainties in the BW effect and calculated electronic HFS factors.

\subsection{Electric quadrupole moments of Po isotopes}

\begin{table*}[]
\caption{Summary of the deduced values of the nuclear magnetic dipole $\mu_I$ and spectroscopic electric quadrupole $Q_S$ moments of Po isotopes.} 
\begin{tabular*}{1.0\textwidth}{l@{\extracolsep{\fill}}lrrrr}
\hline
\hline
Isotope      & I$^{\pi}$  &$\mu_I$ ($\mu_\mathit{N}$)$^a$ & $\mu_I$ ($\mu_\mathit{N}$),& $Q_S$ ($b$)$^b$     & $Q_S$ ($b$),  \\
             &              &~\cite{Seliverstov:2014_Po,Fink:2015,Kowalewska:1991,herzog1983nuclear} & this work & ~\cite{Seliverstov:2014_Po,olsmats1961hyperfine}& this work   \\
\hline
$^{193ls}$Po & (3/2$^-$)  & -0.389(45)     & -0.395(29) & -1.31(34)  & -1.61(16)   \\
$^{193hs}$Po & (13/2$^+$) & -0.742(76)     & -0.753(45) &  1.09(58)  &  1.34(31)  \\
$^{195ls}$Po & (3/2$^-$)  & -0.601(44)     & -0.611(24) & -0.87(29)  & -1.07(16)  \\
$^{195hs}$Po & (13/2$^+$) & -0.932(76)     & -0.946(48) &  1.28(54)  &  1.58(32)  \\
$^{197}$Po   & (3/2$^-$)  & -0.882(66)     & -0.896(32) & -0.44(25)  & -0.54(16)  \\
$^{197m}$Po  & (13/2$^+$) & -1.053(78)     & -1.069(40) &  1.26(54)  &  1.56(31)  \\
$^{199}$Po   & (3/2$^-$)  & -0.912(66)     & -0.926(32) & -0.27(19)  & -0.34(12)  \\
$^{199m}$Po  & (13/2$^+$) & -1.005(81)     & -1.021(50) &  1.40(40)  &  1.72(20)  \\
$^{201}$Po   & 3/2$^-$    & -0.984(71)     & -0.999(37) &  0.10(13)  &  0.118(78)   \\
$^{201m}$Po  & 13/2$^+$   & -1.002(89)     & -1.017(81) &  1.26(40)  &  1.55(20)  \\
$^{203}$Po   & 5/2$^-$    &  0.741(54)     & 0.752(29) &  0.17(13)   &  0.212(78)   \\
$^{203m}$Po  & 13/2$^+$   & -0.965(78)     & -0.980(49) &  1.22(22)  &  1.504(84)   \\
$^{205}$Po   & 5/2$^-$    & 0.760(55)      & 0.775(15)  &  0.14(2)$^c$     & 0.1722(41)   \\
$^{207}$Po   & 5/2$^-$    & 0.793(55)      & 0.805(15)  &  0.23(3)$^{c,d}$ & 0.2859(34)   \\
$^{209}$Po   & 1/2$^-$    & 0.606(45)      & 0.613(18)  & 0          & 0          \\
$^{211}$Po   & 9/2$^+$    & -1.197(90)     & -1.215(50) & -0.77(17)  & -0.940(81)   \\
$^{217}$Po   & (9/2$^+$)  & -1.106(103)    & -1.123(63) &  0.06(44)  &  0.08(31)  \\
\hline
\end{tabular*}
\\
\begin{flushleft}
$^a$ 
Note, that uncertainties in Refs.~\cite{Seliverstov:2014_Po,Fink:2015,Kowalewska:1991} do not include estimation of the possible HFS anomaly.\\
$^b$ Note, that only lower limit of the theoretical uncertainty (10\%) was taken into account for the uncertainty estimation in~\cite{Seliverstov:2014_Po}. Thus, in this column only lower limit of the uncertainties is given.
$^c$ Recalculated in~\cite{Seliverstov:2014_Po} from the value presented in~\cite{olsmats1961hyperfine} taking into account Sternheimer correction.
$^d$ Used in~\cite{Seliverstov:2014_Po} as a reference.
\end{flushleft}
\label{TResultsMoments}  
\end{table*}

The HFS constants $B$ for two electronic states, the ground state $6p^4~^3P_2$ and the excited state $6p^3 7s^1~^5 S_2$, were measured for $^{205}$Po and $^{207}$Po in Refs.~\cite{olsmats1961hyperfine,Kowalewska:1991}. The ratio of the experimental $B$ constants for the two electronic states should be equal to the ratio of the EFG values. The experimental value of the ratio $B(6p^4~^3P_2)/B(6p^3 7s^1~^5 S_2)(^{207}{\rm Po})=1.037(19)$ and our theoretical value $q(6p^4~^3P_2)/q(6p^3 7s^1~^5 S_2)(^{207}{\rm Po})=1.043(25)$ are in good agreement. However, the ratio of the experimental values of the $B$ constants for $^{205}$Po, $B(6p^4~^3P_2)/B(6p^3 7s^1~^5 S_2)(^{205}{\rm Po})=0.951(53)$ deviates by $1.6\sigma$ from both the experimental value for $^{207}{\rm Po}$ and our theoretical result.

Combining the measured values of $B(6p^4~^3P_2)$~\cite{olsmats1961hyperfine} with the calculated EFGs from Table~\ref{TResults}, we obtain:
\begin{eqnarray}
    \label{Q205}
    Q_S(\rm ^{205}{\rm Po}) = 0.1722(41)~b, \\
    \label{Q207}
    Q_S(\rm ^{207}{\rm Po}) = 0.2859(34)~b.
\end{eqnarray}

In Ref.~\cite{Seliverstov:2014_Po}, the value $Q_S(\rm ^{207}{\rm Po})=0.23(3)~b$ was used as a reference to deduce the quadrupole moment of other isotopes. This value was obtained by the semiempirical theoretical approach~\cite{olsmats1961hyperfine} with correction by 20\% to account for the contribution of the core electronic shells to the EFG (Sternheimer correction)~\cite{Sternheimer:1950} since only the valence $6p$ electrons of Po were considered in~\cite{olsmats1961hyperfine} to estimate the values of the EFG. Only lower limit of the theoretical uncertainty was indicated. In the present approach, a rigorous relativistic coupled cluster method was used, and high-order correlation effects for all 84 electrons were explicitly taken into account. To get an idea of the correlation effect of the ``core'' electrons, the value of the EFG for the ground electronic state was calculated at the relativistic CCSD(T) level in two ways: (i) all electrons were included in the correlation treatment, and (ii) only the $6p$ electrons of Po were correlated. By comparing these two calculations, we found that the correlation (which accounts for spatial polarization in particular) of 80 ``core'' electrons of Po increases the value of the EFG by 6\%. A similar calculation at the lower CCSD level gives an increase of 13\%. These estimations qualitatively justify the applied Sterheimer correction, though it was noticeably overestimated (20\% in Ref.~\cite{Seliverstov:2014_Po}). Note that the correlation contribution of the core electrons to the EFG is much smaller than their contribution to the magnetic dipole electronic factor described earlier.

In Refs.~\cite{Kowalewska:1991,Seliverstov:2014_Po,Fink:2015}, HFS constants $B$ were measured for a long chain of Po isotopes for the $6p^3 7s^1~^5 S_2$ excited electronic state. Using our value of the EFG, we reinterpret these measurements. The deduced values of $Q_S$ are given in Table~\ref{TResultsMoments}. Theoretical and experimental uncertainties were summed quadratically.

It is important to stress that all nuclear physics inferences from Refs.~\cite{Seliverstov:2014_Po,Fink:2015} remain valid, though the new $Q_S$ values differ from that used in Refs.~\cite{Seliverstov:2014_Po,Fink:2015} by about 30\%. Moreover, it is the new results that give the firm ground to these inferences, since in fact the previous $Q_S$ values used in~\cite{Seliverstov:2014_Po,Fink:2015} have the uncontrolled theoretical uncertainties. For all isotopes, except $^{207}$Po, the uncertainty is dominated by the experimental uncertainty of the $B$ constant. In the majority of cases, it is about an order of magnitude larger than the uncertainty of the calculated EFG value.

Note, that as in the case of magnetic moments the $6p^4~{}^1D_2$ state proved to be preferable in comparison with the $6p^3 7s^1~^5 S_2$ and $6p^4~^3P_2$ states for the quadrupole moments study. Namely, relative theoretical uncertainty for the EFG($6p^4~{}^1D_2$) is less than that for the EFG($6p^4~^3P_2$) and EFG($6p^3 7s^1~^5 S_2$) by factors of 1.5 and 2.6, respectively. Besides, the HFS splitting and sensitivity to the electromagnetic moments for the $6p^4~{}^1D_2$ state are markedly larger than that for the $6p^3 7s^1~^5 S_2$ and $6p^4~^3P_2$ states (see Table~\ref{TResults}). This means that the components of the HFS spectra for the transition which includes $6p^4~{}^1D_2$ state will be better resolved and as the result we will have lower experimental uncertainties.

\section{Conclusion}
We refined the electronic structure parameters used to interpret the hyperfine structure measurements of neutral polonium. 
In the performed calculations, electronic correlation effects were treated using the coupled cluster method, which includes of up to iterative triple, and perturbative quadruple excitation amplitudes. The Dirac-Coulomb Hamiltonian was used with further correction on the Gaunt interelectron interaction and the effects of quantum electrodynamics. The computational scheme employed enabled us to systematically estimate the uncertainty resulting from various aspects of the calculation, including the completeness of the correlation effects treatment, basis set size, Hamiltonian level, and the nuclear charge distribution effects. We additionally verified the reliability of the proposed approach by comparing the IP, EEs and electronic $g_J$ factor in Po with experiment. The obtained agreement was found to be better than in all previous theoretical studies of these parameters.

Using the available experimental data and our calculated HFS constants we deduced updated values for the nuclear magnetic dipole and electric quadrupole moments of various odd-mass polonium isotopes, significantly reducing the uncertainty (see Table \ref{TResultsMoments}). The reliable values of the quadrupole moments are especially important since in fact the previously used $Q_S$ values had the uncontrolled theoretical uncertainties. Using the parametrization introduced in~\cite{Skripnikov:2020e}, we disentangled contributions from the nuclear magnetization distribution and electronic structure effects: combining calculated parameters with the experimental data for $^{205}$Po and $^{207}$Po we obtained both magnetic moments and their distribution parameters inside the nucleus. The deduced values can be further used to predict hyperfine structure in Po-containing compounds.

We identified the excited electronic state $6p^4~{}^1D_2$ as a promising candidate for further studies on the nuclear moments of polonium. It is expected that the transition involving the $6p^4~{}^1D_2$ state may offer a potentially better experimental resolution and more accurate theoretical interpretation.

 We also made a direct analysis of the contributions to the BW effect in the HFS $A$ constants, specifically due to the spin polarization of the $s_{1/2}$ and $p_{1/2}$ electronic shells described by Eq.~(\ref{ABwexpr}). It was found that in both the ground state and the $6p^3 7s^1~{}^5S_2$ excited state, the BW contribution is strongly dominated by the polarization of the $s_{1/2}$ shells. However, in the $6p^4~{}^1D_2$ state, both the spin polarizations of the $s_{1/2}$ and $p_{1/2}$ shells contribute comparably with opposite signs and with the $p_{1/2}$ contribution being dominant. The partial mutual cancellation of the two contributions leads to a reduced BW effect (and HFS anomaly) for this state. We did not see such behavior in the simpler single-valence electron Ra$^{+}$ ion in its low-lying electronic states~\cite{Skripnikov:2020e}, although a somewhat similar effect was found for the first excited electronic state of the RaF molecule~\cite{Skripnikov:2020e}.

\begin{acknowledgments}  
Electronic structure calculations have been carried out using computing resources of the federal collective usage center Complex for Simulation and Data Processing for Mega-science Facilities at National Research Centre ``Kurchatov Institute'', http://ckp.nrcki.ru/.

The research was supported by the Russian Science Foundation (Grant No. 19-72-10019-P (https://rscf.ru/en/project/22-72-41010/).

\end{acknowledgments}


\begin{thebibliography}{100}%
\makeatletter
\providecommand \@ifxundefined [1]{%
 \@ifx{#1\undefined}
}%
\providecommand \@ifnum [1]{%
 \ifnum #1\expandafter \@firstoftwo
 \else \expandafter \@secondoftwo
 \fi
}%
\providecommand \@ifx [1]{%
 \ifx #1\expandafter \@firstoftwo
 \else \expandafter \@secondoftwo
 \fi
}%
\providecommand \natexlab [1]{#1}%
\providecommand \enquote  [1]{``#1''}%
\providecommand \bibnamefont  [1]{#1}%
\providecommand \bibfnamefont [1]{#1}%
\providecommand \citenamefont [1]{#1}%
\providecommand \href@noop [0]{\@secondoftwo}%
\providecommand \href [0]{\begingroup \@sanitize@url \@href}%
\providecommand \@href[1]{\@@startlink{#1}\@@href}%
\providecommand \@@href[1]{\endgroup#1\@@endlink}%
\providecommand \@sanitize@url [0]{\catcode `\\12\catcode `\$12\catcode
  `\&12\catcode `\#12\catcode `\^12\catcode `\_12\catcode `\%12\relax}%
\providecommand \@@startlink[1]{}%
\providecommand \@@endlink[0]{}%
\providecommand \url  [0]{\begingroup\@sanitize@url \@url }%
\providecommand \@url [1]{\endgroup\@href {#1}{\urlprefix }}%
\providecommand \urlprefix  [0]{URL }%
\providecommand \Eprint [0]{\href }%
\providecommand \doibase [0]{http://dx.doi.org/}%
\providecommand \selectlanguage [0]{\@gobble}%
\providecommand \bibinfo  [0]{\@secondoftwo}%
\providecommand \bibfield  [0]{\@secondoftwo}%
\providecommand \translation [1]{[#1]}%
\providecommand \BibitemOpen [0]{}%
\providecommand \bibitemStop [0]{}%
\providecommand \bibitemNoStop [0]{.\EOS\space}%
\providecommand \EOS [0]{\spacefactor3000\relax}%
\providecommand \BibitemShut  [1]{\csname bibitem#1\endcsname}%
\let\auto@bib@innerbib\@empty
\bibitem [{\citenamefont {Yang}\ \emph {et~al.}(2022)\citenamefont {Yang},
  \citenamefont {Wang}, \citenamefont {Wilkins},\ and\ \citenamefont
  {Garcia~Ruiz}}]{Yang2022_exotic}%
  \BibitemOpen
  \bibfield  {author} {\bibinfo {author} {\bibfnamefont {X.~F.}\ \bibnamefont
  {Yang}}, \bibinfo {author} {\bibfnamefont {S.~J.}\ \bibnamefont {Wang}},
  \bibinfo {author} {\bibfnamefont {S.~G.}\ \bibnamefont {Wilkins}}, \ and\
  \bibinfo {author} {\bibfnamefont {R.~F.}\ \bibnamefont {Garcia~Ruiz}},\
  }\href {\doibase https://doi.org/10.1016/j.ppnp.2022.104005} {\bibfield
  {journal} {\bibinfo  {journal} {Prog. Part. Nucl. Phys.}\ ,\ \bibinfo {pages}
  {104005}} (\bibinfo {year} {2022})}\BibitemShut {NoStop}%
\bibitem [{\citenamefont {Sassarini}\ \emph {et~al.}(2022)\citenamefont
  {Sassarini}, \citenamefont {Dobaczewski}, \citenamefont {Bonnard},\ and\
  \citenamefont {Ruiz}}]{Sassarini:2022}%
  \BibitemOpen
  \bibfield  {author} {\bibinfo {author} {\bibfnamefont {P.~L.}\ \bibnamefont
  {Sassarini}}, \bibinfo {author} {\bibfnamefont {J.}~\bibnamefont
  {Dobaczewski}}, \bibinfo {author} {\bibfnamefont {J.}~\bibnamefont
  {Bonnard}}, \ and\ \bibinfo {author} {\bibfnamefont {R.~F.~G.}\ \bibnamefont
  {Ruiz}},\ }\href {\doibase 10.1088/1361-6471/ac900a} {\bibfield  {journal}
  {\bibinfo  {journal} {J. Phys. G: Nucl. Part. Phys.}\ }\textbf {\bibinfo
  {volume} {49}},\ \bibinfo {pages} {11LT01} (\bibinfo {year}
  {2022})}\BibitemShut {NoStop}%
\bibitem [{\citenamefont {Bonnard}\ \emph {et~al.}(2023)\citenamefont
  {Bonnard}, \citenamefont {Dobaczewski}, \citenamefont {Danneaux},\ and\
  \citenamefont {Kortelainen}}]{NucDFT:2023}%
  \BibitemOpen
  \bibfield  {author} {\bibinfo {author} {\bibfnamefont {J.}~\bibnamefont
  {Bonnard}}, \bibinfo {author} {\bibfnamefont {J.}~\bibnamefont
  {Dobaczewski}}, \bibinfo {author} {\bibfnamefont {G.}~\bibnamefont
  {Danneaux}}, \ and\ \bibinfo {author} {\bibfnamefont {M.}~\bibnamefont
  {Kortelainen}},\ }\href {\doibase   https://doi.org/10.1016/j.physletb.2023.138014} {\bibfield  {journal}
  {\bibinfo  {journal} {Phys. Lett. B}\ }\textbf {\bibinfo {volume} {843}},\
  \bibinfo {pages} {138014} (\bibinfo {year} {2023})}\BibitemShut {NoStop}%
\bibitem [{\citenamefont {Shabaev}\ \emph
  {et~al.}(2001{\natexlab{a}})\citenamefont {Shabaev}, \citenamefont
  {Artemyev}, \citenamefont {Yerokhin}, \citenamefont {Zherebtsov},\ and\
  \citenamefont {Soff}}]{Shabaev:01a}%
  \BibitemOpen
  \bibfield  {author} {\bibinfo {author} {\bibfnamefont {V.~M.}\ \bibnamefont
  {Shabaev}}, \bibinfo {author} {\bibfnamefont {A.~N.}\ \bibnamefont
  {Artemyev}}, \bibinfo {author} {\bibfnamefont {V.~A.}\ \bibnamefont
  {Yerokhin}}, \bibinfo {author} {\bibfnamefont {O.~M.}\ \bibnamefont
  {Zherebtsov}}, \ and\ \bibinfo {author} {\bibfnamefont {G.}~\bibnamefont
  {Soff}},\ }\href {\doibase 10.1103/PhysRevLett.86.3959} {\bibfield  {journal}
  {\bibinfo  {journal} {Phys.\ Rev.\ Lett.}\ }\textbf {\bibinfo {volume}
  {86}},\ \bibinfo {pages} {3959} (\bibinfo {year}
  {2001}{\natexlab{a}})}\BibitemShut {NoStop}%
\bibitem [{\citenamefont {Skripnikov}\ \emph {et~al.}(2018)\citenamefont
  {Skripnikov}, \citenamefont {Schmidt}, \citenamefont {Ullmann}, \citenamefont
  {Geppert}, \citenamefont {Kraus}, \citenamefont {Kresse}, \citenamefont
  {N\"ortersh\"auser}, \citenamefont {Privalov}, \citenamefont {Scheibe},
  \citenamefont {Shabaev}, \citenamefont {Vogel},\ and\ \citenamefont
  {Volotka}}]{Skripnikov:18a}%
  \BibitemOpen
  \bibfield  {author} {\bibinfo {author} {\bibfnamefont {L.~V.}\ \bibnamefont
  {Skripnikov}}, \bibinfo {author} {\bibfnamefont {S.}~\bibnamefont {Schmidt}},
  \bibinfo {author} {\bibfnamefont {J.}~\bibnamefont {Ullmann}}, \bibinfo
  {author} {\bibfnamefont {C.}~\bibnamefont {Geppert}}, \bibinfo {author}
  {\bibfnamefont {F.}~\bibnamefont {Kraus}}, \bibinfo {author} {\bibfnamefont
  {B.}~\bibnamefont {Kresse}}, \bibinfo {author} {\bibfnamefont
  {W.}~\bibnamefont {N\"ortersh\"auser}}, \bibinfo {author} {\bibfnamefont
  {A.~F.}\ \bibnamefont {Privalov}}, \bibinfo {author} {\bibfnamefont
  {B.}~\bibnamefont {Scheibe}}, \bibinfo {author} {\bibfnamefont {V.~M.}\
  \bibnamefont {Shabaev}}, \bibinfo {author} {\bibfnamefont {M.}~\bibnamefont
  {Vogel}}, \ and\ \bibinfo {author} {\bibfnamefont {A.~V.}\ \bibnamefont
  {Volotka}},\ }\href {\doibase 10.1103/PhysRevLett.120.093001} {\bibfield
  {journal} {\bibinfo  {journal} {Phys.\ Rev.\ Lett.}\ }\textbf {\bibinfo
  {volume} {120}},\ \bibinfo {pages} {093001} (\bibinfo {year}
  {2018})}\BibitemShut {NoStop}%
\bibitem [{\citenamefont {N{\"o}rtersh{\"a}user}\ \emph
  {et~al.}(2019)\citenamefont {N{\"o}rtersh{\"a}user}, \citenamefont {Ullmann},
  \citenamefont {Skripnikov}, \citenamefont {Andelkovic}, \citenamefont
  {Brandau}, \citenamefont {Dax}, \citenamefont {Geithner}, \citenamefont
  {Geppert}, \citenamefont {Gorges}, \citenamefont {Hammen}, \citenamefont
  {Hannen}, \citenamefont {Kaufmann}, \citenamefont {K{\"o}nig}, \citenamefont
  {Kraus}, \citenamefont {Kresse}, \citenamefont {Litvinov}, \citenamefont
  {Lochmann}, \citenamefont {Maa{\ss}}, \citenamefont {Meisner}, \citenamefont
  {Murb{\"o}ck}, \citenamefont {Privalov}, \citenamefont {S{\'a}nchez},
  \citenamefont {Scheibe}, \citenamefont {Schmidt}, \citenamefont {Schmidt},
  \citenamefont {Shabaev}, \citenamefont {Steck}, \citenamefont {St{\"o}hlker},
  \citenamefont {Thompson}, \citenamefont {Trageser}, \citenamefont {Vogel},
  \citenamefont {Vollbrecht}, \citenamefont {Volotka},\ and\ \citenamefont
  {Weinheimer}}]{Nortershauser:2019}%
  \BibitemOpen
  \bibfield  {author} {\bibinfo {author} {\bibfnamefont {W.}~\bibnamefont
  {N{\"o}rtersh{\"a}user}}, \bibinfo {author} {\bibfnamefont {J.}~\bibnamefont
  {Ullmann}}, \bibinfo {author} {\bibfnamefont {L.~V.}\ \bibnamefont
  {Skripnikov}}, \bibinfo {author} {\bibfnamefont {Z.}~\bibnamefont
  {Andelkovic}}, \bibinfo {author} {\bibfnamefont {C.}~\bibnamefont {Brandau}},
  \bibinfo {author} {\bibfnamefont {A.}~\bibnamefont {Dax}}, \bibinfo {author}
  {\bibfnamefont {W.}~\bibnamefont {Geithner}}, \bibinfo {author}
  {\bibfnamefont {C.}~\bibnamefont {Geppert}}, \bibinfo {author} {\bibfnamefont
  {C.}~\bibnamefont {Gorges}}, \bibinfo {author} {\bibfnamefont
  {M.}~\bibnamefont {Hammen}}, \bibinfo {author} {\bibfnamefont
  {V.}~\bibnamefont {Hannen}}, \bibinfo {author} {\bibfnamefont
  {S.}~\bibnamefont {Kaufmann}}, \bibinfo {author} {\bibfnamefont
  {K.}~\bibnamefont {K{\"o}nig}}, \bibinfo {author} {\bibfnamefont
  {F.}~\bibnamefont {Kraus}}, \bibinfo {author} {\bibfnamefont
  {B.}~\bibnamefont {Kresse}}, \bibinfo {author} {\bibfnamefont {Y.~A.}\
  \bibnamefont {Litvinov}}, \bibinfo {author} {\bibfnamefont {M.}~\bibnamefont
  {Lochmann}}, \bibinfo {author} {\bibfnamefont {B.}~\bibnamefont {Maa{\ss}}},
  \bibinfo {author} {\bibfnamefont {J.}~\bibnamefont {Meisner}}, \bibinfo
  {author} {\bibfnamefont {T.}~\bibnamefont {Murb{\"o}ck}}, \bibinfo {author}
  {\bibfnamefont {A.~F.}\ \bibnamefont {Privalov}}, \bibinfo {author}
  {\bibfnamefont {R.}~\bibnamefont {S{\'a}nchez}}, \bibinfo {author}
  {\bibfnamefont {B.}~\bibnamefont {Scheibe}}, \bibinfo {author} {\bibfnamefont
  {M.}~\bibnamefont {Schmidt}}, \bibinfo {author} {\bibfnamefont
  {S.}~\bibnamefont {Schmidt}}, \bibinfo {author} {\bibfnamefont {V.~M.}\
  \bibnamefont {Shabaev}}, \bibinfo {author} {\bibfnamefont {M.}~\bibnamefont
  {Steck}}, \bibinfo {author} {\bibfnamefont {T.}~\bibnamefont {St{\"o}hlker}},
  \bibinfo {author} {\bibfnamefont {R.~C.}\ \bibnamefont {Thompson}}, \bibinfo
  {author} {\bibfnamefont {C.}~\bibnamefont {Trageser}}, \bibinfo {author}
  {\bibfnamefont {M.}~\bibnamefont {Vogel}}, \bibinfo {author} {\bibfnamefont
  {J.}~\bibnamefont {Vollbrecht}}, \bibinfo {author} {\bibfnamefont {A.~V.}\
  \bibnamefont {Volotka}}, \ and\ \bibinfo {author} {\bibfnamefont
  {C.}~\bibnamefont {Weinheimer}},\ }\href {\doibase 10.1007/s10751-019-1569-8}
  {\bibfield  {journal} {\bibinfo  {journal} {Hyperfine Interact.}\ }\textbf
  {\bibinfo {volume} {240}},\ \bibinfo {pages} {51} (\bibinfo {year}
  {2019})}\BibitemShut {NoStop}%
\bibitem [{\citenamefont {Skripnikov}\ and\ \citenamefont
  {Prosnyak}(2022)}]{Skripnikov:2022}%
  \BibitemOpen
  \bibfield  {author} {\bibinfo {author} {\bibfnamefont {L.~V.}\ \bibnamefont
  {Skripnikov}}\ and\ \bibinfo {author} {\bibfnamefont {S.~D.}\ \bibnamefont
  {Prosnyak}},\ }\href {\doibase 10.1103/PhysRevC.106.054303} {\bibfield
  {journal} {\bibinfo  {journal} {Phys. Rev. C}\ }\textbf {\bibinfo {volume}
  {106}},\ \bibinfo {pages} {054303} (\bibinfo {year} {2022})}\BibitemShut
  {NoStop}%
\bibitem [{\citenamefont {Fella}\ \emph {et~al.}(2020)\citenamefont {Fella},
  \citenamefont {Skripnikov}, \citenamefont {N\"ortersh\"auser}, \citenamefont
  {Buchner}, \citenamefont {Deubner}, \citenamefont {Kraus}, \citenamefont
  {Privalov}, \citenamefont {Shabaev},\ and\ \citenamefont
  {Vogel}}]{Skripnikov:2020a}%
  \BibitemOpen
  \bibfield  {author} {\bibinfo {author} {\bibfnamefont {V.}~\bibnamefont
  {Fella}}, \bibinfo {author} {\bibfnamefont {L.~V.}\ \bibnamefont
  {Skripnikov}}, \bibinfo {author} {\bibfnamefont {W.}~\bibnamefont
  {N\"ortersh\"auser}}, \bibinfo {author} {\bibfnamefont {M.~R.}\ \bibnamefont
  {Buchner}}, \bibinfo {author} {\bibfnamefont {H.~L.}\ \bibnamefont
  {Deubner}}, \bibinfo {author} {\bibfnamefont {F.}~\bibnamefont {Kraus}},
  \bibinfo {author} {\bibfnamefont {A.~F.}\ \bibnamefont {Privalov}}, \bibinfo
  {author} {\bibfnamefont {V.~M.}\ \bibnamefont {Shabaev}}, \ and\ \bibinfo
  {author} {\bibfnamefont {M.}~\bibnamefont {Vogel}},\ }\href {\doibase   10.1103/PhysRevResearch.2.013368} {\bibfield  {journal} {\bibinfo  {journal}
  {Phys. Rev. Research}\ }\textbf {\bibinfo {volume} {2}},\ \bibinfo {pages}
  {013368} (\bibinfo {year} {2020})}\BibitemShut {NoStop}%
\bibitem [{\citenamefont {Pyykk{\"o}}(2008)}]{Pekka:2008}%
  \BibitemOpen
  \bibfield  {author} {\bibinfo {author} {\bibfnamefont {P.}~\bibnamefont
  {Pyykk{\"o}}},\ }\href {\doibase 10.1080/00268970802018367} {\bibfield
  {journal} {\bibinfo  {journal} {Mol. Phys.}\ }\textbf {\bibinfo {volume}
  {106}},\ \bibinfo {pages} {1965} (\bibinfo {year} {2008})}\BibitemShut
  {NoStop}%
\bibitem [{\citenamefont {Kozlov}(1997)}]{Kozlov:97c}%
  \BibitemOpen
  \bibfield  {author} {\bibinfo {author} {\bibfnamefont {M.~G.}\ \bibnamefont
  {Kozlov}},\ }\href {\doibase 10.1088/0953-4075/30/18/003} {\bibfield
  {journal} {\bibinfo  {journal} {J.\ Phys.\ B}\ }\textbf {\bibinfo {volume}
  {30}},\ \bibinfo {pages} {L607} (\bibinfo {year} {1997})}\BibitemShut
  {NoStop}%
\bibitem [{\citenamefont {Safronova}\ \emph {et~al.}(2018)\citenamefont
  {Safronova}, \citenamefont {Budker}, \citenamefont {DeMille}, \citenamefont
  {Kimball}, \citenamefont {Derevianko},\ and\ \citenamefont
  {Clark}}]{Safronova:18}%
  \BibitemOpen
  \bibfield  {author} {\bibinfo {author} {\bibfnamefont {M.~S.}\ \bibnamefont
  {Safronova}}, \bibinfo {author} {\bibfnamefont {D.}~\bibnamefont {Budker}},
  \bibinfo {author} {\bibfnamefont {D.}~\bibnamefont {DeMille}}, \bibinfo
  {author} {\bibfnamefont {D.~F.~J.}\ \bibnamefont {Kimball}}, \bibinfo
  {author} {\bibfnamefont {A.}~\bibnamefont {Derevianko}}, \ and\ \bibinfo
  {author} {\bibfnamefont {C.~W.}\ \bibnamefont {Clark}},\ }\href {\doibase   10.1103/RevModPhys.90.025008} {\bibfield  {journal} {\bibinfo  {journal}
  {Rev.\ Mod.\ Phys.}\ }\textbf {\bibinfo {volume} {90}},\ \bibinfo {pages}
  {025008} (\bibinfo {year} {2018})}\BibitemShut {NoStop}%
\bibitem [{\citenamefont {Porsev}\ \emph {et~al.}(2009)\citenamefont {Porsev},
  \citenamefont {Beloy},\ and\ \citenamefont {Derevianko}}]{Porsev:2009}%
  \BibitemOpen
  \bibfield  {author} {\bibinfo {author} {\bibfnamefont {S.~G.}\ \bibnamefont
  {Porsev}}, \bibinfo {author} {\bibfnamefont {K.}~\bibnamefont {Beloy}}, \
  and\ \bibinfo {author} {\bibfnamefont {A.}~\bibnamefont {Derevianko}},\
  }\href {\doibase 10.1103/PhysRevLett.102.181601} {\bibfield  {journal}
  {\bibinfo  {journal} {Phys. Rev. Lett.}\ }\textbf {\bibinfo {volume} {102}},\
  \bibinfo {pages} {181601} (\bibinfo {year} {2009})}\BibitemShut {NoStop}%
\bibitem [{\citenamefont {Ginges}\ \emph {et~al.}(2017)\citenamefont {Ginges},
  \citenamefont {Volotka},\ and\ \citenamefont {Fritzsche}}]{ginges2017ground}%
  \BibitemOpen
  \bibfield  {author} {\bibinfo {author} {\bibfnamefont {J.~S.~M.}\
  \bibnamefont {Ginges}}, \bibinfo {author} {\bibfnamefont {A.~V.}\
  \bibnamefont {Volotka}}, \ and\ \bibinfo {author} {\bibfnamefont
  {S.}~\bibnamefont {Fritzsche}},\ }\href {\doibase 10.1103/PhysRevA.96.062502}
  {\bibfield  {journal} {\bibinfo  {journal} {Phys.\ Rev.\ A}\ }\textbf
  {\bibinfo {volume} {96}},\ \bibinfo {pages} {062502} (\bibinfo {year}
  {2017})}\BibitemShut {NoStop}%
\bibitem [{\citenamefont {Fleig}\ and\ \citenamefont
  {Skripnikov}(2020)}]{Skripnikov:2020b}%
  \BibitemOpen
  \bibfield  {author} {\bibinfo {author} {\bibfnamefont {T.}~\bibnamefont
  {Fleig}}\ and\ \bibinfo {author} {\bibfnamefont {L.~V.}\ \bibnamefont
  {Skripnikov}},\ }\href {\doibase 10.3390/sym12040498} {\bibfield  {journal}
  {\bibinfo  {journal} {Symmetry}\ }\textbf {\bibinfo {volume} {12}},\ \bibinfo
  {pages} {498} (\bibinfo {year} {2020})}\BibitemShut {NoStop}%
\bibitem [{\citenamefont {Ginges}\ and\ \citenamefont
  {Flambaum}(2004)}]{GFreview}%
  \BibitemOpen
  \bibfield  {author} {\bibinfo {author} {\bibfnamefont {J.~S.~M.}\
  \bibnamefont {Ginges}}\ and\ \bibinfo {author} {\bibfnamefont {V.~V.}\
  \bibnamefont {Flambaum}},\ }\href {\doibase 10.1016/j.physrep.2004.03.005}
  {\bibfield  {journal} {\bibinfo  {journal} {Phys.\ Rep.}\ }\textbf {\bibinfo
  {volume} {397}},\ \bibinfo {pages} {63} (\bibinfo {year} {2004})}\BibitemShut
  {NoStop}%
\bibitem [{\citenamefont {Kozlov}\ and\ \citenamefont
  {Labzowsky}(1995)}]{KL95}%
  \BibitemOpen
  \bibfield  {author} {\bibinfo {author} {\bibfnamefont {M.}~\bibnamefont
  {Kozlov}}\ and\ \bibinfo {author} {\bibfnamefont {L.}~\bibnamefont
  {Labzowsky}},\ }\href {\doibase 10.1088/0953-4075/28/10/008} {\bibfield
  {journal} {\bibinfo  {journal} {J.\ Phys.\ B}\ }\textbf {\bibinfo {volume}
  {28}},\ \bibinfo {pages} {1933} (\bibinfo {year} {1995})}\BibitemShut
  {NoStop}%
\bibitem [{\citenamefont {Quiney}\ \emph {et~al.}(1998)\citenamefont {Quiney},
  \citenamefont {Skaane},\ and\ \citenamefont {Grant}}]{Quiney:98}%
  \BibitemOpen
  \bibfield  {author} {\bibinfo {author} {\bibfnamefont {H.~M.}\ \bibnamefont
  {Quiney}}, \bibinfo {author} {\bibfnamefont {H.}~\bibnamefont {Skaane}}, \
  and\ \bibinfo {author} {\bibfnamefont {I.~P.}\ \bibnamefont {Grant}},\ }\href
  {\doibase 10.1088/0953-4075/31/3/003} {\bibfield  {journal} {\bibinfo
  {journal} {J.\ Phys.\ B}\ }\textbf {\bibinfo {volume} {31}},\ \bibinfo
  {pages} {L85} (\bibinfo {year} {1998})}\BibitemShut {NoStop}%
\bibitem [{\citenamefont {Titov}\ \emph {et~al.}(2006)\citenamefont {Titov},
  \citenamefont {Mosyagin}, \citenamefont {Petrov}, \citenamefont {Isaev},\
  and\ \citenamefont {DeMille}}]{Titov:06amin}%
  \BibitemOpen
  \bibfield  {author} {\bibinfo {author} {\bibfnamefont {A.~V.}\ \bibnamefont
  {Titov}}, \bibinfo {author} {\bibfnamefont {N.~S.}\ \bibnamefont {Mosyagin}},
  \bibinfo {author} {\bibfnamefont {A.~N.}\ \bibnamefont {Petrov}}, \bibinfo
  {author} {\bibfnamefont {T.~A.}\ \bibnamefont {Isaev}}, \ and\ \bibinfo
  {author} {\bibfnamefont {D.~P.}\ \bibnamefont {DeMille}},\ }\href {\doibase
  10.1007/1-4020-4528-X_12} {\bibfield  {journal} {\bibinfo  {journal} {Progr.\
  Theor.\ Chem.\ Phys.}\ }\textbf {\bibinfo {volume} {15}},\ \bibinfo {pages}
  {253} (\bibinfo {year} {2006})}\BibitemShut {NoStop}%
\bibitem [{\citenamefont {Skripnikov}\ and\ \citenamefont
  {Titov}(2015{\natexlab{a}})}]{Skripnikov:15b}%
  \BibitemOpen
  \bibfield  {author} {\bibinfo {author} {\bibfnamefont {L.~V.}\ \bibnamefont
  {Skripnikov}}\ and\ \bibinfo {author} {\bibfnamefont {A.~V.}\ \bibnamefont
  {Titov}},\ }\href {\doibase 10.1103/PhysRevA.91.042504} {\bibfield  {journal}
  {\bibinfo  {journal} {Phys. Rev. A}\ }\textbf {\bibinfo {volume} {91}},\
  \bibinfo {pages} {042504} (\bibinfo {year} {2015}{\natexlab{a}})}\BibitemShut
  {NoStop}%
\bibitem [{\citenamefont {Skripnikov}\ and\ \citenamefont
  {Titov}(2015{\natexlab{b}})}]{Skripnikov:15a}%
  \BibitemOpen
  \bibfield  {author} {\bibinfo {author} {\bibfnamefont {L.~V.}\ \bibnamefont
  {Skripnikov}}\ and\ \bibinfo {author} {\bibfnamefont {A.~V.}\ \bibnamefont
  {Titov}},\ }\href {\doibase 10.1063/1.4904877} {\bibfield  {journal}
  {\bibinfo  {journal} {J.\ Chem.\ Phys.}\ }\textbf {\bibinfo {volume} {142}},\
  \bibinfo {eid} {024301} (\bibinfo {year} {2015}{\natexlab{b}})}\BibitemShut
  {NoStop}%
\bibitem [{\citenamefont {Sunaga}\ \emph {et~al.}(2016)\citenamefont {Sunaga},
  \citenamefont {Abe}, \citenamefont {Hada},\ and\ \citenamefont
  {Das}}]{Sunaga:16}%
  \BibitemOpen
  \bibfield  {author} {\bibinfo {author} {\bibfnamefont {A.}~\bibnamefont
  {Sunaga}}, \bibinfo {author} {\bibfnamefont {M.}~\bibnamefont {Abe}},
  \bibinfo {author} {\bibfnamefont {M.}~\bibnamefont {Hada}}, \ and\ \bibinfo
  {author} {\bibfnamefont {B.~P.}\ \bibnamefont {Das}},\ }\href {\doibase   10.1103/PhysRevA.93.042507} {\bibfield  {journal} {\bibinfo  {journal}
  {Phys.\ Rev.\ A}\ }\textbf {\bibinfo {volume} {93}},\ \bibinfo {pages}
  {042507} (\bibinfo {year} {2016})}\BibitemShut {NoStop}%
\bibitem [{\citenamefont {Fleig}(2017)}]{Fleig:17}%
  \BibitemOpen
  \bibfield  {author} {\bibinfo {author} {\bibfnamefont {T.}~\bibnamefont
  {Fleig}},\ }\href {\doibase 10.1103/PhysRevA.96.040502} {\bibfield  {journal}
  {\bibinfo  {journal} {Phys.\ Rev.\ A}\ }\textbf {\bibinfo {volume} {96}},\
  \bibinfo {pages} {040502(R)} (\bibinfo {year} {2017})}\BibitemShut {NoStop}%
\bibitem [{\citenamefont {Haase}\ \emph {et~al.}(2020)\citenamefont {Haase},
  \citenamefont {Eliav}, \citenamefont {Ilia\v{s}},\ and\ \citenamefont
  {Borschevsky}}]{Borschevsky:2020}%
  \BibitemOpen
  \bibfield  {author} {\bibinfo {author} {\bibfnamefont {P.~A.~B.}\
  \bibnamefont {Haase}}, \bibinfo {author} {\bibfnamefont {E.}~\bibnamefont
  {Eliav}}, \bibinfo {author} {\bibfnamefont {M.}~\bibnamefont {Ilia\v{s}}}, \
  and\ \bibinfo {author} {\bibfnamefont {A.}~\bibnamefont {Borschevsky}},\
  }\href {\doibase 10.1021/acs.jpca.0c00877} {\bibfield  {journal} {\bibinfo
  {journal} {J. Phys. Chem. A}\ }\textbf {\bibinfo {volume} {124}},\ \bibinfo
  {pages} {3157} (\bibinfo {year} {2020})}\BibitemShut {NoStop}%
\bibitem [{\citenamefont {Skripnikov}(2020)}]{Skripnikov:2020e}%
  \BibitemOpen
  \bibfield  {author} {\bibinfo {author} {\bibfnamefont {L.~V.}\ \bibnamefont
  {Skripnikov}},\ }\href {\doibase 10.1063/5.0024103} {\bibfield  {journal}
  {\bibinfo  {journal} {J.\ Chem.\ Phys.}\ }\textbf {\bibinfo {volume} {153}},\
  \bibinfo {pages} {114114} (\bibinfo {year} {2020})}\BibitemShut {NoStop}%
\bibitem [{\citenamefont {Seliverstov}\ \emph {et~al.}(2014)\citenamefont
  {Seliverstov}, \citenamefont {Cocolios}, \citenamefont {Dexters},
  \citenamefont {Andreyev}, \citenamefont {Antalic}, \citenamefont {Barzakh},
  \citenamefont {Bastin}, \citenamefont {B\"uscher}, \citenamefont {Darby},
  \citenamefont {Fedorov}, \citenamefont {Fedosseev}, \citenamefont {Flanagan},
  \citenamefont {Franchoo}, \citenamefont {Huber}, \citenamefont {Huyse},
  \citenamefont {Keupers}, \citenamefont {K\"oster}, \citenamefont
  {Kudryavtsev}, \citenamefont {Marsh}, \citenamefont {Molkanov}, \citenamefont
  {Page}, \citenamefont {Sj\"odin}, \citenamefont {Stefan}, \citenamefont
  {Van~Duppen}, \citenamefont {Venhart},\ and\ \citenamefont
  {Zemlyanoy}}]{Seliverstov:2014_Po}%
  \BibitemOpen
  \bibfield  {author} {\bibinfo {author} {\bibfnamefont {M.~D.}\ \bibnamefont
  {Seliverstov}}, \bibinfo {author} {\bibfnamefont {T.~E.}\ \bibnamefont
  {Cocolios}}, \bibinfo {author} {\bibfnamefont {W.}~\bibnamefont {Dexters}},
  \bibinfo {author} {\bibfnamefont {A.~N.}\ \bibnamefont {Andreyev}}, \bibinfo
  {author} {\bibfnamefont {S.}~\bibnamefont {Antalic}}, \bibinfo {author}
  {\bibfnamefont {A.~E.}\ \bibnamefont {Barzakh}}, \bibinfo {author}
  {\bibfnamefont {B.}~\bibnamefont {Bastin}}, \bibinfo {author} {\bibfnamefont
  {J.}~\bibnamefont {B\"uscher}}, \bibinfo {author} {\bibfnamefont {I.~G.}\
  \bibnamefont {Darby}}, \bibinfo {author} {\bibfnamefont {D.~V.}\ \bibnamefont
  {Fedorov}}, \bibinfo {author} {\bibfnamefont {V.~N.}\ \bibnamefont
  {Fedosseev}}, \bibinfo {author} {\bibfnamefont {K.~T.}\ \bibnamefont
  {Flanagan}}, \bibinfo {author} {\bibfnamefont {S.}~\bibnamefont {Franchoo}},
  \bibinfo {author} {\bibfnamefont {G.}~\bibnamefont {Huber}}, \bibinfo
  {author} {\bibfnamefont {M.}~\bibnamefont {Huyse}}, \bibinfo {author}
  {\bibfnamefont {M.}~\bibnamefont {Keupers}}, \bibinfo {author} {\bibfnamefont
  {U.}~\bibnamefont {K\"oster}}, \bibinfo {author} {\bibfnamefont
  {Y.}~\bibnamefont {Kudryavtsev}}, \bibinfo {author} {\bibfnamefont {B.~A.}\
  \bibnamefont {Marsh}}, \bibinfo {author} {\bibfnamefont {P.~L.}\ \bibnamefont
  {Molkanov}}, \bibinfo {author} {\bibfnamefont {R.~D.}\ \bibnamefont {Page}},
  \bibinfo {author} {\bibfnamefont {A.~M.}\ \bibnamefont {Sj\"odin}}, \bibinfo
  {author} {\bibfnamefont {I.}~\bibnamefont {Stefan}}, \bibinfo {author}
  {\bibfnamefont {P.}~\bibnamefont {Van~Duppen}}, \bibinfo {author}
  {\bibfnamefont {M.}~\bibnamefont {Venhart}}, \ and\ \bibinfo {author}
  {\bibfnamefont {S.~G.}\ \bibnamefont {Zemlyanoy}},\ }\href {\doibase   10.1103/PhysRevC.89.034323} {\bibfield  {journal} {\bibinfo  {journal} {Phys.
  Rev. C}\ }\textbf {\bibinfo {volume} {89}},\ \bibinfo {pages} {034323}
  (\bibinfo {year} {2014})}\BibitemShut {NoStop}%
\bibitem [{\citenamefont {Olsmats}\ \emph {et~al.}(1961)\citenamefont
  {Olsmats}, \citenamefont {Axensten},\ and\ \citenamefont
  {Liljegren}}]{olsmats1961hyperfine}%
  \BibitemOpen
  \bibfield  {author} {\bibinfo {author} {\bibfnamefont {C.}~\bibnamefont
  {Olsmats}}, \bibinfo {author} {\bibfnamefont {S.}~\bibnamefont {Axensten}}, \
  and\ \bibinfo {author} {\bibfnamefont {G.}~\bibnamefont {Liljegren}},\
  }\href@noop {} {\bibfield  {journal} {\bibinfo  {journal} {Arkiv f\"or
  Fysik}\ }\textbf {\bibinfo {volume} {19}},\ \bibinfo {pages} {469} (\bibinfo
  {year} {1961})}\BibitemShut {NoStop}%
\bibitem [{\citenamefont {Kowalewska}\ \emph {et~al.}(1991)\citenamefont
  {Kowalewska}, \citenamefont {Bekk}, \citenamefont {G\"oring}, \citenamefont
  {Hanser}, \citenamefont {K\"alber}, \citenamefont {Meisel},\ and\
  \citenamefont {Rebel}}]{Kowalewska:1991}%
  \BibitemOpen
  \bibfield  {author} {\bibinfo {author} {\bibfnamefont {D.}~\bibnamefont
  {Kowalewska}}, \bibinfo {author} {\bibfnamefont {K.}~\bibnamefont {Bekk}},
  \bibinfo {author} {\bibfnamefont {S.}~\bibnamefont {G\"oring}}, \bibinfo
  {author} {\bibfnamefont {A.}~\bibnamefont {Hanser}}, \bibinfo {author}
  {\bibfnamefont {W.}~\bibnamefont {K\"alber}}, \bibinfo {author}
  {\bibfnamefont {G.}~\bibnamefont {Meisel}}, \ and\ \bibinfo {author}
  {\bibfnamefont {H.}~\bibnamefont {Rebel}},\ }\href {\doibase   10.1103/PhysRevA.44.R1442} {\bibfield  {journal} {\bibinfo  {journal} {Phys.
  Rev. A}\ }\textbf {\bibinfo {volume} {44}},\ \bibinfo {pages} {R1442}
  (\bibinfo {year} {1991})}\BibitemShut {NoStop}%
\bibitem [{\citenamefont {Fink}\ \emph {et~al.}(2015)\citenamefont {Fink},
  \citenamefont {Cocolios}, \citenamefont {Andreyev}, \citenamefont {Antalic},
  \citenamefont {Barzakh}, \citenamefont {Bastin}, \citenamefont {Fedorov},
  \citenamefont {Fedosseev}, \citenamefont {Flanagan}, \citenamefont {Ghys},
  \citenamefont {Gottberg}, \citenamefont {Huyse}, \citenamefont {Imai},
  \citenamefont {Kron}, \citenamefont {Lecesne}, \citenamefont {Lynch},
  \citenamefont {Marsh}, \citenamefont {Pauwels}, \citenamefont {Rapisarda},
  \citenamefont {Richter}, \citenamefont {Rossel}, \citenamefont {Rothe},
  \citenamefont {Seliverstov}, \citenamefont {Sj\"odin}, \citenamefont
  {Van~Beveren}, \citenamefont {Van~Duppen},\ and\ \citenamefont
  {Wendt}}]{Fink:2015}%
  \BibitemOpen
  \bibfield  {author} {\bibinfo {author} {\bibfnamefont {D.~A.}\ \bibnamefont
  {Fink}}, \bibinfo {author} {\bibfnamefont {T.~E.}\ \bibnamefont {Cocolios}},
  \bibinfo {author} {\bibfnamefont {A.~N.}\ \bibnamefont {Andreyev}}, \bibinfo
  {author} {\bibfnamefont {S.}~\bibnamefont {Antalic}}, \bibinfo {author}
  {\bibfnamefont {A.~E.}\ \bibnamefont {Barzakh}}, \bibinfo {author}
  {\bibfnamefont {B.}~\bibnamefont {Bastin}}, \bibinfo {author} {\bibfnamefont
  {D.~V.}\ \bibnamefont {Fedorov}}, \bibinfo {author} {\bibfnamefont {V.~N.}\
  \bibnamefont {Fedosseev}}, \bibinfo {author} {\bibfnamefont {K.~T.}\
  \bibnamefont {Flanagan}}, \bibinfo {author} {\bibfnamefont {L.}~\bibnamefont
  {Ghys}}, \bibinfo {author} {\bibfnamefont {A.}~\bibnamefont {Gottberg}},
  \bibinfo {author} {\bibfnamefont {M.}~\bibnamefont {Huyse}}, \bibinfo
  {author} {\bibfnamefont {N.}~\bibnamefont {Imai}}, \bibinfo {author}
  {\bibfnamefont {T.}~\bibnamefont {Kron}}, \bibinfo {author} {\bibfnamefont
  {N.}~\bibnamefont {Lecesne}}, \bibinfo {author} {\bibfnamefont {K.~M.}\
  \bibnamefont {Lynch}}, \bibinfo {author} {\bibfnamefont {B.~A.}\ \bibnamefont
  {Marsh}}, \bibinfo {author} {\bibfnamefont {D.}~\bibnamefont {Pauwels}},
  \bibinfo {author} {\bibfnamefont {E.}~\bibnamefont {Rapisarda}}, \bibinfo
  {author} {\bibfnamefont {S.~D.}\ \bibnamefont {Richter}}, \bibinfo {author}
  {\bibfnamefont {R.~E.}\ \bibnamefont {Rossel}}, \bibinfo {author}
  {\bibfnamefont {S.}~\bibnamefont {Rothe}}, \bibinfo {author} {\bibfnamefont
  {M.~D.}\ \bibnamefont {Seliverstov}}, \bibinfo {author} {\bibfnamefont
  {A.~M.}\ \bibnamefont {Sj\"odin}}, \bibinfo {author} {\bibfnamefont
  {C.}~\bibnamefont {Van~Beveren}}, \bibinfo {author} {\bibfnamefont
  {P.}~\bibnamefont {Van~Duppen}}, \ and\ \bibinfo {author} {\bibfnamefont
  {K.~D.~A.}\ \bibnamefont {Wendt}},\ }\href {\doibase   10.1103/PhysRevX.5.011018} {\bibfield  {journal} {\bibinfo  {journal} {Phys.
  Rev. X}\ }\textbf {\bibinfo {volume} {5}},\ \bibinfo {pages} {011018}
  (\bibinfo {year} {2015})}\BibitemShut {NoStop}%
\bibitem [{\citenamefont {Finkelnburg}\ and\ \citenamefont
  {Stern}(1950)}]{Finkelnburg:1950}%
  \BibitemOpen
  \bibfield  {author} {\bibinfo {author} {\bibfnamefont {W.}~\bibnamefont
  {Finkelnburg}}\ and\ \bibinfo {author} {\bibfnamefont {F.}~\bibnamefont
  {Stern}},\ }\href {\doibase 10.1103/PhysRev.77.303.2} {\bibfield  {journal}
  {\bibinfo  {journal} {Phys. Rev.}\ }\textbf {\bibinfo {volume} {77}},\
  \bibinfo {pages} {303} (\bibinfo {year} {1950})}\BibitemShut {NoStop}%
\bibitem [{\citenamefont {Charles}\ \emph {et~al.}(1955)\citenamefont
  {Charles}, \citenamefont {Hunt}, \citenamefont {Pish},\ and\ \citenamefont
  {Timma}}]{Charles:55}%
  \BibitemOpen
  \bibfield  {author} {\bibinfo {author} {\bibfnamefont {G.~W.}\ \bibnamefont
  {Charles}}, \bibinfo {author} {\bibfnamefont {D.~J.}\ \bibnamefont {Hunt}},
  \bibinfo {author} {\bibfnamefont {G.}~\bibnamefont {Pish}}, \ and\ \bibinfo
  {author} {\bibfnamefont {D.~L.}\ \bibnamefont {Timma}},\ }\href {\doibase  10.1364/JOSA.45.000869} {\bibfield  {journal} {\bibinfo  {journal} {J. Opt.
  Soc. Am.}\ }\textbf {\bibinfo {volume} {45}},\ \bibinfo {pages} {869}
  (\bibinfo {year} {1955})}\BibitemShut {NoStop}%
\bibitem [{\citenamefont {Charles}(1966)}]{Charles:66}%
  \BibitemOpen
  \bibfield  {author} {\bibinfo {author} {\bibfnamefont {G.~W.}\ \bibnamefont
  {Charles}},\ }\href {\doibase 10.1364/JOSA.56.001292} {\bibfield  {journal}
  {\bibinfo  {journal} {J. Opt. Soc. Am.}\ }\textbf {\bibinfo {volume} {56}},\
  \bibinfo {pages} {1292} (\bibinfo {year} {1966})}\BibitemShut {NoStop}%
\bibitem [{\citenamefont {Raeder}\ \emph {et~al.}(2019)\citenamefont {Raeder},
  \citenamefont {Heggen}, \citenamefont {Teigelhöfer},\ and\ \citenamefont
  {Lassen}}]{Raeder:2019}%
  \BibitemOpen
  \bibfield  {author} {\bibinfo {author} {\bibfnamefont {S.}~\bibnamefont
  {Raeder}}, \bibinfo {author} {\bibfnamefont {H.}~\bibnamefont {Heggen}},
  \bibinfo {author} {\bibfnamefont {A.}~\bibnamefont {Teigelhöfer}}, \ and\
  \bibinfo {author} {\bibfnamefont {J.}~\bibnamefont {Lassen}},\ }\href
  {\doibase https://doi.org/10.1016/j.sab.2018.08.005} {\bibfield  {journal}
  {\bibinfo  {journal} {Spectrochim. Acta Part B At. Spectrosc.}\ }\textbf
  {\bibinfo {volume} {151}},\ \bibinfo {pages} {65} (\bibinfo {year}
  {2019})}\BibitemShut {NoStop}%
\bibitem [{\citenamefont {Fink}\ \emph {et~al.}(2019)\citenamefont {Fink},
  \citenamefont {Blaum}, \citenamefont {Fedosseev}, \citenamefont {Marsh},
  \citenamefont {Rossel},\ and\ \citenamefont {Rothe}}]{Po_IPexp:2019}%
  \BibitemOpen
  \bibfield  {author} {\bibinfo {author} {\bibfnamefont {D.}~\bibnamefont
  {Fink}}, \bibinfo {author} {\bibfnamefont {K.}~\bibnamefont {Blaum}},
  \bibinfo {author} {\bibfnamefont {V.}~\bibnamefont {Fedosseev}}, \bibinfo
  {author} {\bibfnamefont {B.}~\bibnamefont {Marsh}}, \bibinfo {author}
  {\bibfnamefont {R.}~\bibnamefont {Rossel}}, \ and\ \bibinfo {author}
  {\bibfnamefont {S.}~\bibnamefont {Rothe}},\ }\href {\doibase   https://doi.org/10.1016/j.sab.2018.08.004} {\bibfield  {journal} {\bibinfo
  {journal} {Spectrochim. Acta Part B At. Spectrosc.}\ }\textbf {\bibinfo
  {volume} {151}},\ \bibinfo {pages} {72} (\bibinfo {year} {2019})}\BibitemShut
  {NoStop}%
\bibitem [{\citenamefont {Axensten}\ and\ \citenamefont
  {Olsmats}(1961)}]{axensten1961nuclear}%
  \BibitemOpen
  \bibfield  {author} {\bibinfo {author} {\bibfnamefont {S.}~\bibnamefont
  {Axensten}}\ and\ \bibinfo {author} {\bibfnamefont {C.}~\bibnamefont
  {Olsmats}},\ }\href@noop {} {\bibfield  {journal} {\bibinfo  {journal} {Arkiv
  Fysik}\ }\textbf {\bibinfo {volume} {19}},\ \bibinfo {pages} {461} (\bibinfo
  {year} {1961})}\BibitemShut {NoStop}%
\bibitem [{\citenamefont {Herzog}\ \emph {et~al.}(1983)\citenamefont {Herzog},
  \citenamefont {Walitzki}, \citenamefont {Freitag}, \citenamefont
  {Hildebrand},\ and\ \citenamefont {Schl{\"o}sser}}]{herzog1983nuclear}%
  \BibitemOpen
  \bibfield  {author} {\bibinfo {author} {\bibfnamefont {P.}~\bibnamefont
  {Herzog}}, \bibinfo {author} {\bibfnamefont {H.}~\bibnamefont {Walitzki}},
  \bibinfo {author} {\bibfnamefont {K.}~\bibnamefont {Freitag}}, \bibinfo
  {author} {\bibfnamefont {H.}~\bibnamefont {Hildebrand}}, \ and\ \bibinfo
  {author} {\bibfnamefont {K.}~\bibnamefont {Schl{\"o}sser}},\ }\href {\doibase   doi.org/10.1007/BF01415691} {\bibfield  {journal} {\bibinfo  {journal}
  {Zeitschrift f{\"u}r Physik A Atoms and Nuclei}\ }\textbf {\bibinfo {volume}
  {311}},\ \bibinfo {pages} {351} (\bibinfo {year} {1983})}\BibitemShut
  {NoStop}%
\bibitem [{\citenamefont {Rosenthal}\ and\ \citenamefont
  {Breit}(1932)}]{rosenthal1932isotope}%
  \BibitemOpen
  \bibfield  {author} {\bibinfo {author} {\bibfnamefont {J.~E.}\ \bibnamefont
  {Rosenthal}}\ and\ \bibinfo {author} {\bibfnamefont {G.}~\bibnamefont
  {Breit}},\ }\href {\doibase 10.1103/PhysRev.41.459} {\bibfield  {journal}
  {\bibinfo  {journal} {Phys.\ Rev.}\ }\textbf {\bibinfo {volume} {41}},\
  \bibinfo {pages} {459} (\bibinfo {year} {1932})}\BibitemShut {NoStop}%
\bibitem [{\citenamefont {Crawford}\ and\ \citenamefont
  {Schawlow}(1949)}]{crawford1949mf}%
  \BibitemOpen
  \bibfield  {author} {\bibinfo {author} {\bibfnamefont {M.~F.}\ \bibnamefont
  {Crawford}}\ and\ \bibinfo {author} {\bibfnamefont {A.~L.}\ \bibnamefont
  {Schawlow}},\ }\href {\doibase 10.1103/PhysRev.76.1310} {\bibfield  {journal}
  {\bibinfo  {journal} {Phys.\ Rev.}\ }\textbf {\bibinfo {volume} {76}},\
  \bibinfo {pages} {1310} (\bibinfo {year} {1949})}\BibitemShut {NoStop}%
\bibitem [{\citenamefont {Bohr}\ and\ \citenamefont
  {Weisskopf}(1950)}]{bohr1950influence}%
  \BibitemOpen
  \bibfield  {author} {\bibinfo {author} {\bibfnamefont {A.}~\bibnamefont
  {Bohr}}\ and\ \bibinfo {author} {\bibfnamefont {V.~F.}\ \bibnamefont
  {Weisskopf}},\ }\href {\doibase 10.1103/PhysRev.77.94} {\bibfield  {journal}
  {\bibinfo  {journal} {Phys.\ Rev.}\ }\textbf {\bibinfo {volume} {77}},\
  \bibinfo {pages} {94} (\bibinfo {year} {1950})}\BibitemShut {NoStop}%
\bibitem [{\citenamefont {Bohr}(1951)}]{bohr1951bohr}%
  \BibitemOpen
  \bibfield  {author} {\bibinfo {author} {\bibfnamefont {A.}~\bibnamefont
  {Bohr}},\ }\href {\doibase 10.1103/PhysRev.81.134} {\bibfield  {journal}
  {\bibinfo  {journal} {Phys. Rev.}\ }\textbf {\bibinfo {volume} {81}},\
  \bibinfo {pages} {134} (\bibinfo {year} {1951})}\BibitemShut {NoStop}%
\bibitem [{\citenamefont {Sliv}(1951)}]{sliv1951uchet}%
  \BibitemOpen
  \bibfield  {author} {\bibinfo {author} {\bibfnamefont {L.}~\bibnamefont
  {Sliv}},\ }\href@noop {} {\bibfield  {journal} {\bibinfo  {journal} {Zh.
  Eksp. Teor. Fiz.}\ }\textbf {\bibinfo {volume} {21}},\ \bibinfo {pages} {770}
  (\bibinfo {year} {1951})}\BibitemShut {NoStop}%
\bibitem [{\citenamefont {Johnson}(2007)}]{johnson2007atomic}%
  \BibitemOpen
  \bibfield  {author} {\bibinfo {author} {\bibfnamefont {W.~R.}\ \bibnamefont
  {Johnson}},\ }\href@noop {} {\emph {\bibinfo {title} {Atomic structure
  theory}}}\ (\bibinfo  {publisher} {Springer},\ \bibinfo {year}
  {2007})\BibitemShut {NoStop}%
\bibitem [{\citenamefont {Zherebtsov}\ and\ \citenamefont
  {Shabaev}(2000)}]{Zherebtsov:2000}%
  \BibitemOpen
  \bibfield  {author} {\bibinfo {author} {\bibfnamefont {O.~M.}\ \bibnamefont
  {Zherebtsov}}\ and\ \bibinfo {author} {\bibfnamefont {V.~M.}\ \bibnamefont
  {Shabaev}},\ }\href {\doibase 10.1139/p00-060} {\bibfield  {journal}
  {\bibinfo  {journal} {Can. J. Phys.}\ }\textbf {\bibinfo {volume} {78}},\
  \bibinfo {pages} {701} (\bibinfo {year} {2000})}\BibitemShut {NoStop}%
\bibitem [{\citenamefont {Tupitsyn}\ \emph {et~al.}(2002)\citenamefont
  {Tupitsyn}, \citenamefont {Loginov},\ and\ \citenamefont
  {Shabaev}}]{Tupitsyn:02}%
  \BibitemOpen
  \bibfield  {author} {\bibinfo {author} {\bibfnamefont {I.~I.}\ \bibnamefont
  {Tupitsyn}}, \bibinfo {author} {\bibfnamefont {A.~V.}\ \bibnamefont
  {Loginov}}, \ and\ \bibinfo {author} {\bibfnamefont {V.~M.}\ \bibnamefont
  {Shabaev}},\ }\href {\doibase 10.1134/1.1509815} {\bibfield  {journal}
  {\bibinfo  {journal} {Opt. Spectrosc.}\ }\textbf {\bibinfo {volume} {93}},\
  \bibinfo {pages} {357} (\bibinfo {year} {2002})}\BibitemShut {NoStop}%
\bibitem [{\citenamefont {Malkin}\ \emph {et~al.}(2011)\citenamefont {Malkin},
  \citenamefont {Repisky}, \citenamefont {Komorovsky}, \citenamefont {Mach},
  \citenamefont {Malkina},\ and\ \citenamefont {Malkin}}]{Malkin:2011}%
  \BibitemOpen
  \bibfield  {author} {\bibinfo {author} {\bibfnamefont {E.}~\bibnamefont
  {Malkin}}, \bibinfo {author} {\bibfnamefont {M.}~\bibnamefont {Repisky}},
  \bibinfo {author} {\bibfnamefont {S.}~\bibnamefont {Komorovsky}}, \bibinfo
  {author} {\bibfnamefont {P.}~\bibnamefont {Mach}}, \bibinfo {author}
  {\bibfnamefont {O.~L.}\ \bibnamefont {Malkina}}, \ and\ \bibinfo {author}
  {\bibfnamefont {V.~G.}\ \bibnamefont {Malkin}},\ }\href {\doibase
  10.1063/1.3526263} {\bibfield  {journal} {\bibinfo  {journal} {J.\ Chem.\
  Phys.}\ }\textbf {\bibinfo {volume} {134}},\ \bibinfo {pages} {044111}
  (\bibinfo {year} {2011})}\BibitemShut {NoStop}%
\bibitem [{\citenamefont {Schwartz}(1955)}]{Schwartz:1955}%
  \BibitemOpen
  \bibfield  {author} {\bibinfo {author} {\bibfnamefont {C.}~\bibnamefont
  {Schwartz}},\ }\href {\doibase 10.1103/PhysRev.99.1035} {\bibfield  {journal}
  {\bibinfo  {journal} {Phys. Rev.}\ }\textbf {\bibinfo {volume} {99}},\
  \bibinfo {pages} {1035} (\bibinfo {year} {1955})}\BibitemShut {NoStop}%
\bibitem [{\citenamefont {Persson}(1998)}]{Persson1998}%
  \BibitemOpen
  \bibfield  {author} {\bibinfo {author} {\bibfnamefont {J.}~\bibnamefont
  {Persson}},\ }\href {\doibase 10.1007/s100500050081} {\bibfield  {journal}
  {\bibinfo  {journal} {Eur.\ Phys.\ J.\ A}\ }\textbf {\bibinfo {volume} {2}},\
  \bibinfo {pages} {3} (\bibinfo {year} {1998})}\BibitemShut {NoStop}%
\bibitem [{\citenamefont {Prosnyak}\ and\ \citenamefont
  {Skripnikov}(2021)}]{Prosnyak:2021}%
  \BibitemOpen
  \bibfield  {author} {\bibinfo {author} {\bibfnamefont {S.~D.}\ \bibnamefont
  {Prosnyak}}\ and\ \bibinfo {author} {\bibfnamefont {L.~V.}\ \bibnamefont
  {Skripnikov}},\ }\href {\doibase 10.1103/PhysRevC.103.034314} {\bibfield
  {journal} {\bibinfo  {journal} {Phys. Rev. C}\ }\textbf {\bibinfo {volume}
  {103}},\ \bibinfo {pages} {034314} (\bibinfo {year} {2021})}\BibitemShut
  {NoStop}%
\bibitem [{\citenamefont {Konovalova}\ \emph {et~al.}(2017)\citenamefont
  {Konovalova}, \citenamefont {Kozlov}, \citenamefont {Demidov},\ and\
  \citenamefont {Barzakh}}]{konovalova2017calculation}%
  \BibitemOpen
  \bibfield  {author} {\bibinfo {author} {\bibfnamefont {E.~A.}\ \bibnamefont
  {Konovalova}}, \bibinfo {author} {\bibfnamefont {M.~G.}\ \bibnamefont
  {Kozlov}}, \bibinfo {author} {\bibfnamefont {Y.~A.}\ \bibnamefont {Demidov}},
  \ and\ \bibinfo {author} {\bibfnamefont {A.~E.}\ \bibnamefont {Barzakh}},\
  }\href {arXiv:1703.10048} {\bibfield  {journal} {\bibinfo  {journal} {Rad.
  Appl.}\ }\textbf {\bibinfo {volume} {2}},\ \bibinfo {pages} {181} (\bibinfo
  {year} {2017})}\BibitemShut {NoStop}%
\bibitem [{\citenamefont {Konovalova}\ \emph {et~al.}(2018)\citenamefont
  {Konovalova}, \citenamefont {Demidov}, \citenamefont {Kozlov},\ and\
  \citenamefont {Barzakh}}]{atoms6030039}%
  \BibitemOpen
  \bibfield  {author} {\bibinfo {author} {\bibfnamefont {E.~A.}\ \bibnamefont
  {Konovalova}}, \bibinfo {author} {\bibfnamefont {Y.~A.}\ \bibnamefont
  {Demidov}}, \bibinfo {author} {\bibfnamefont {M.~G.}\ \bibnamefont {Kozlov}},
  \ and\ \bibinfo {author} {\bibfnamefont {A.~E.}\ \bibnamefont {Barzakh}},\
  }\href {\doibase 10.3390/atoms6030039} {\bibfield  {journal} {\bibinfo
  {journal} {Atoms}\ }\textbf {\bibinfo {volume} {6}},\ \bibinfo {pages} {39}
  (\bibinfo {year} {2018})}\BibitemShut {NoStop}%
\bibitem [{\citenamefont {Shabaev}\ \emph {et~al.}(1997)\citenamefont
  {Shabaev}, \citenamefont {Tomaselli}, \citenamefont {K{\"u}hl}, \citenamefont
  {Artemyev},\ and\ \citenamefont {Yerokhin}}]{shabaev1997ground}%
  \BibitemOpen
  \bibfield  {author} {\bibinfo {author} {\bibfnamefont {V.~M.}\ \bibnamefont
  {Shabaev}}, \bibinfo {author} {\bibfnamefont {M.}~\bibnamefont {Tomaselli}},
  \bibinfo {author} {\bibfnamefont {T.}~\bibnamefont {K{\"u}hl}}, \bibinfo
  {author} {\bibfnamefont {A.~N.}\ \bibnamefont {Artemyev}}, \ and\ \bibinfo
  {author} {\bibfnamefont {V.~A.}\ \bibnamefont {Yerokhin}},\ }\href {\doibase   10.1103/PhysRevA.56.252} {\bibfield  {journal} {\bibinfo  {journal} {Phys.\
  Rev.\ A}\ }\textbf {\bibinfo {volume} {56}},\ \bibinfo {pages} {252}
  (\bibinfo {year} {1997})}\BibitemShut {NoStop}%
\bibitem [{\citenamefont {Shabaev}\ \emph
  {et~al.}(2001{\natexlab{b}})\citenamefont {Shabaev}, \citenamefont
  {Yerokhin}, \citenamefont {Zherebtsov}, \citenamefont {Artemyev},
  \citenamefont {Sysak},\ and\ \citenamefont {Soff}}]{Shabaev:01}%
  \BibitemOpen
  \bibfield  {author} {\bibinfo {author} {\bibfnamefont {V.~M.}\ \bibnamefont
  {Shabaev}}, \bibinfo {author} {\bibfnamefont {V.~A.}\ \bibnamefont
  {Yerokhin}}, \bibinfo {author} {\bibfnamefont {O.~M.}\ \bibnamefont
  {Zherebtsov}}, \bibinfo {author} {\bibfnamefont {A.~N.}\ \bibnamefont
  {Artemyev}}, \bibinfo {author} {\bibfnamefont {M.~M.}\ \bibnamefont {Sysak}},
  \ and\ \bibinfo {author} {\bibfnamefont {G.}~\bibnamefont {Soff}},\ }\href
  {\doibase 10.1023/A:1011961031356} {\bibfield  {journal} {\bibinfo  {journal}
  {Hyperfine Interact.}\ }\textbf {\bibinfo {volume} {132}},\ \bibinfo {pages}
  {341} (\bibinfo {year} {2001}{\natexlab{b}})}\BibitemShut {NoStop}%
\bibitem [{\citenamefont {Volotka}\ \emph {et~al.}(2008)\citenamefont
  {Volotka}, \citenamefont {Glazov}, \citenamefont {Tupitsyn}, \citenamefont
  {Oreshkina}, \citenamefont {Plunien},\ and\ \citenamefont
  {Shabaev}}]{Volotka:2008}%
  \BibitemOpen
  \bibfield  {author} {\bibinfo {author} {\bibfnamefont {A.~V.}\ \bibnamefont
  {Volotka}}, \bibinfo {author} {\bibfnamefont {D.~A.}\ \bibnamefont {Glazov}},
  \bibinfo {author} {\bibfnamefont {I.~I.}\ \bibnamefont {Tupitsyn}}, \bibinfo
  {author} {\bibfnamefont {N.~S.}\ \bibnamefont {Oreshkina}}, \bibinfo {author}
  {\bibfnamefont {G.}~\bibnamefont {Plunien}}, \ and\ \bibinfo {author}
  {\bibfnamefont {V.~M.}\ \bibnamefont {Shabaev}},\ }\href {\doibase   10.1103/PhysRevA.78.062507} {\bibfield  {journal} {\bibinfo  {journal} {Phys.
  Rev. A}\ }\textbf {\bibinfo {volume} {78}},\ \bibinfo {pages} {062507}
  (\bibinfo {year} {2008})}\BibitemShut {NoStop}%
\bibitem [{\citenamefont {Porsev}\ \emph {et~al.}(2021)\citenamefont {Porsev},
  \citenamefont {Safronova},\ and\ \citenamefont {Kozlov}}]{Porsev:2021}%
  \BibitemOpen
  \bibfield  {author} {\bibinfo {author} {\bibfnamefont {S.~G.}\ \bibnamefont
  {Porsev}}, \bibinfo {author} {\bibfnamefont {M.~S.}\ \bibnamefont
  {Safronova}}, \ and\ \bibinfo {author} {\bibfnamefont {M.~G.}\ \bibnamefont
  {Kozlov}},\ }\href {\doibase 10.1103/PhysRevLett.127.253001} {\bibfield
  {journal} {\bibinfo  {journal} {Phys. Rev. Lett.}\ }\textbf {\bibinfo
  {volume} {127}},\ \bibinfo {pages} {253001} (\bibinfo {year}
  {2021})}\BibitemShut {NoStop}%
\bibitem [{\citenamefont {Ehlers}\ \emph {et~al.}(1968)\citenamefont {Ehlers},
  \citenamefont {Kabasakal}, \citenamefont {Shugart},\ and\ \citenamefont
  {Tezer}}]{Ehlers:1968}%
  \BibitemOpen
  \bibfield  {author} {\bibinfo {author} {\bibfnamefont {V.~J.}\ \bibnamefont
  {Ehlers}}, \bibinfo {author} {\bibfnamefont {Y.}~\bibnamefont {Kabasakal}},
  \bibinfo {author} {\bibfnamefont {H.~A.}\ \bibnamefont {Shugart}}, \ and\
  \bibinfo {author} {\bibfnamefont {O.}~\bibnamefont {Tezer}},\ }\href
  {\doibase 10.1103/PhysRev.176.25} {\bibfield  {journal} {\bibinfo  {journal}
  {Phys. Rev.}\ }\textbf {\bibinfo {volume} {176}},\ \bibinfo {pages} {25}
  (\bibinfo {year} {1968})}\BibitemShut {NoStop}%
\bibitem [{\citenamefont {Barzakh}\ \emph {et~al.}(2012)\citenamefont
  {Barzakh}, \citenamefont {Batist}, \citenamefont {Fedorov}, \citenamefont
  {Ivanov}, \citenamefont {Mezilev}, \citenamefont {Molkanov}, \citenamefont
  {Moroz}, \citenamefont {Orlov}, \citenamefont {Panteleev},\ and\
  \citenamefont {Volkov}}]{barzakh2012hyperfine}%
  \BibitemOpen
  \bibfield  {author} {\bibinfo {author} {\bibfnamefont {A.~E.}\ \bibnamefont
  {Barzakh}}, \bibinfo {author} {\bibfnamefont {L.~K.}\ \bibnamefont {Batist}},
  \bibinfo {author} {\bibfnamefont {D.~V.}\ \bibnamefont {Fedorov}}, \bibinfo
  {author} {\bibfnamefont {V.~S.}\ \bibnamefont {Ivanov}}, \bibinfo {author}
  {\bibfnamefont {K.~A.}\ \bibnamefont {Mezilev}}, \bibinfo {author}
  {\bibfnamefont {P.~L.}\ \bibnamefont {Molkanov}}, \bibinfo {author}
  {\bibfnamefont {F.~V.}\ \bibnamefont {Moroz}}, \bibinfo {author}
  {\bibfnamefont {S.~Y.}\ \bibnamefont {Orlov}}, \bibinfo {author}
  {\bibfnamefont {V.~N.}\ \bibnamefont {Panteleev}}, \ and\ \bibinfo {author}
  {\bibfnamefont {Y.~M.}\ \bibnamefont {Volkov}},\ }\href {\doibase   10.1103/PhysRevC.86.014311} {\bibfield  {journal} {\bibinfo  {journal}
  {Phys.\ Rev.\ C}\ }\textbf {\bibinfo {volume} {86}},\ \bibinfo {pages}
  {014311} (\bibinfo {year} {2012})}\BibitemShut {NoStop}%
\bibitem [{\citenamefont {Schmidt}\ \emph {et~al.}(2018)\citenamefont
  {Schmidt}, \citenamefont {Billowes}, \citenamefont {Bissell}, \citenamefont
  {Blaum}, \citenamefont {Ruiz}, \citenamefont {Heylen}, \citenamefont
  {Malbrunot-Ettenauer}, \citenamefont {Neyens}, \citenamefont
  {N{\"o}rtersh{\"a}user}, \citenamefont {Plunien}, \citenamefont {Sailer},
  \citenamefont {Shabaev}, \citenamefont {Skripnikov}, \citenamefont
  {Tupitsyn}, \citenamefont {Volotka},\ and\ \citenamefont
  {Yang}}]{Schmidt:2018}%
  \BibitemOpen
  \bibfield  {author} {\bibinfo {author} {\bibfnamefont {S.}~\bibnamefont
  {Schmidt}}, \bibinfo {author} {\bibfnamefont {J.}~\bibnamefont {Billowes}},
  \bibinfo {author} {\bibfnamefont {M.~L.}\ \bibnamefont {Bissell}}, \bibinfo
  {author} {\bibfnamefont {K.}~\bibnamefont {Blaum}}, \bibinfo {author}
  {\bibfnamefont {R.~F.~G.}\ \bibnamefont {Ruiz}}, \bibinfo {author}
  {\bibfnamefont {H.}~\bibnamefont {Heylen}}, \bibinfo {author} {\bibfnamefont
  {S.}~\bibnamefont {Malbrunot-Ettenauer}}, \bibinfo {author} {\bibfnamefont
  {G.}~\bibnamefont {Neyens}}, \bibinfo {author} {\bibfnamefont
  {W.}~\bibnamefont {N{\"o}rtersh{\"a}user}}, \bibinfo {author} {\bibfnamefont
  {G.}~\bibnamefont {Plunien}}, \bibinfo {author} {\bibfnamefont
  {S.}~\bibnamefont {Sailer}}, \bibinfo {author} {\bibfnamefont {V.~M.}\
  \bibnamefont {Shabaev}}, \bibinfo {author} {\bibfnamefont {L.~V.}\
  \bibnamefont {Skripnikov}}, \bibinfo {author} {\bibfnamefont {I.~I.}\
  \bibnamefont {Tupitsyn}}, \bibinfo {author} {\bibfnamefont {A.~V.}\
  \bibnamefont {Volotka}}, \ and\ \bibinfo {author} {\bibfnamefont {X.~F.}\
  \bibnamefont {Yang}},\ }\href {\doibase   https://doi.org/10.1016/j.physletb.2018.02.024} {\bibfield  {journal}
  {\bibinfo  {journal} {Phys.\ Lett.\ B}\ }\textbf {\bibinfo {volume} {779}},\
  \bibinfo {pages} {324 } (\bibinfo {year} {2018})}\BibitemShut {NoStop}%
\bibitem [{\citenamefont {Roberts}\ \emph {et~al.}(2022)\citenamefont
  {Roberts}, \citenamefont {Ranclaud},\ and\ \citenamefont
  {Ginges}}]{Ginges:2022}%
  \BibitemOpen
  \bibfield  {author} {\bibinfo {author} {\bibfnamefont {B.~M.}\ \bibnamefont
  {Roberts}}, \bibinfo {author} {\bibfnamefont {P.~G.}\ \bibnamefont
  {Ranclaud}}, \ and\ \bibinfo {author} {\bibfnamefont {J.~S.~M.}\ \bibnamefont
  {Ginges}},\ }\href {\doibase 10.1103/PhysRevA.105.052802} {\bibfield
  {journal} {\bibinfo  {journal} {Phys. Rev. A}\ }\textbf {\bibinfo {volume}
  {105}},\ \bibinfo {pages} {052802} (\bibinfo {year} {2022})}\BibitemShut
  {NoStop}%
\bibitem [{\citenamefont {Cheng}\ and\ \citenamefont
  {Childs}(1985)}]{Cheng:85}%
  \BibitemOpen
  \bibfield  {author} {\bibinfo {author} {\bibfnamefont {K.~T.}\ \bibnamefont
  {Cheng}}\ and\ \bibinfo {author} {\bibfnamefont {W.~J.}\ \bibnamefont
  {Childs}},\ }\href {\doibase 10.1103/PhysRevA.31.2775} {\bibfield  {journal}
  {\bibinfo  {journal} {Phys.\ Rev.\ A}\ }\textbf {\bibinfo {volume} {31}},\
  \bibinfo {pages} {2775} (\bibinfo {year} {1985})}\BibitemShut {NoStop}%
\bibitem [{\citenamefont {Visscher}\ \emph {et~al.}(1996)\citenamefont
  {Visscher}, \citenamefont {Lee},\ and\ \citenamefont {Dyall}}]{Visscher:96a}%
  \BibitemOpen
  \bibfield  {author} {\bibinfo {author} {\bibfnamefont {L.}~\bibnamefont
  {Visscher}}, \bibinfo {author} {\bibfnamefont {T.~J.}\ \bibnamefont {Lee}}, \
  and\ \bibinfo {author} {\bibfnamefont {K.~G.}\ \bibnamefont {Dyall}},\ }\href
  {\doibase 10.1063/1.472655} {\bibfield  {journal} {\bibinfo  {journal} {J.\
  Chem.\ Phys.}\ }\textbf {\bibinfo {volume} {105}},\ \bibinfo {pages} {8769}
  (\bibinfo {year} {1996})}\BibitemShut {NoStop}%
\bibitem [{\citenamefont {Bartlett}\ and\ \citenamefont
  {Musia{\l}}(2007)}]{Bartlett:2007}%
  \BibitemOpen
  \bibfield  {author} {\bibinfo {author} {\bibfnamefont {R.~J.}\ \bibnamefont
  {Bartlett}}\ and\ \bibinfo {author} {\bibfnamefont {M.}~\bibnamefont
  {Musia{\l}}},\ }\href {\doibase 10.1103/RevModPhys.79.291} {\bibfield
  {journal} {\bibinfo  {journal} {Rev. Mod. Phys.}\ }\textbf {\bibinfo {volume}
  {79}},\ \bibinfo {pages} {291} (\bibinfo {year} {2007})}\BibitemShut
  {NoStop}%
\bibitem [{\citenamefont {Skripnikov}\ \emph {et~al.}(2017)\citenamefont
  {Skripnikov}, \citenamefont {Maison},\ and\ \citenamefont
  {Mosyagin}}]{Skripnikov:17a}%
  \BibitemOpen
  \bibfield  {author} {\bibinfo {author} {\bibfnamefont {L.~V.}\ \bibnamefont
  {Skripnikov}}, \bibinfo {author} {\bibfnamefont {D.~E.}\ \bibnamefont
  {Maison}}, \ and\ \bibinfo {author} {\bibfnamefont {N.~S.}\ \bibnamefont
  {Mosyagin}},\ }\href {\doibase 10.1103/PhysRevA.95.022507} {\bibfield
  {journal} {\bibinfo  {journal} {Phys.\ Rev.\ A}\ }\textbf {\bibinfo {volume}
  {95}},\ \bibinfo {pages} {022507} (\bibinfo {year} {2017})}\BibitemShut
  {NoStop}%
\bibitem [{\citenamefont {Dyall}(2006)}]{Dyall:06}%
  \BibitemOpen
  \bibfield  {author} {\bibinfo {author} {\bibfnamefont {K.~G.}\ \bibnamefont
  {Dyall}},\ }\href {\doibase 10.1007/s00214-006-0126-0} {\bibfield  {journal}
  {\bibinfo  {journal} {Theor. Chem. Acc.}\ }\textbf {\bibinfo {volume}
  {115}},\ \bibinfo {pages} {441} (\bibinfo {year} {2006})}\BibitemShut
  {NoStop}%
\bibitem [{\citenamefont {Dyall}(2012)}]{Dyall:12}%
  \BibitemOpen
  \bibfield  {author} {\bibinfo {author} {\bibfnamefont {K.~G.}\ \bibnamefont
  {Dyall}},\ }\href {\doibase 10.1007/s00214-012-1217-8} {\bibfield  {journal}
  {\bibinfo  {journal} {Theor. Chem. Acc.}\ }\textbf {\bibinfo {volume}
  {131}},\ \bibinfo {pages} {1217} (\bibinfo {year} {2012})}\BibitemShut
  {NoStop}%
\bibitem [{\citenamefont {Noga}\ \emph {et~al.}(1987)\citenamefont {Noga},
  \citenamefont {Bartlett},\ and\ \citenamefont {Urban}}]{Noga:CCSDT-3:87}%
  \BibitemOpen
  \bibfield  {author} {\bibinfo {author} {\bibfnamefont {J.}~\bibnamefont
  {Noga}}, \bibinfo {author} {\bibfnamefont {R.~J.}\ \bibnamefont {Bartlett}},
  \ and\ \bibinfo {author} {\bibfnamefont {M.}~\bibnamefont {Urban}},\ }\href
  {\doibase 10.1016/0009-2614(87)87107-5} {\bibfield  {journal} {\bibinfo
  {journal} {Chem. Phys. Lett.}\ }\textbf {\bibinfo {volume} {134}},\ \bibinfo
  {pages} {126} (\bibinfo {year} {1987})}\BibitemShut {NoStop}%
\bibitem [{\citenamefont {Skripnikov}\ \emph {et~al.}(2013)\citenamefont
  {Skripnikov}, \citenamefont {Mosyagin},\ and\ \citenamefont
  {Titov}}]{Skripnikov:13a}%
  \BibitemOpen
  \bibfield  {author} {\bibinfo {author} {\bibfnamefont {L.~V.}\ \bibnamefont
  {Skripnikov}}, \bibinfo {author} {\bibfnamefont {N.~S.}\ \bibnamefont
  {Mosyagin}}, \ and\ \bibinfo {author} {\bibfnamefont {A.~V.}\ \bibnamefont
  {Titov}},\ }\href {\doibase 10.1016/j.cplett.2012.11.013} {\bibfield
  {journal} {\bibinfo  {journal} {Chem.\ Phys.\ Lett.}\ }\textbf {\bibinfo
  {volume} {555}},\ \bibinfo {pages} {79} (\bibinfo {year} {2013})}\BibitemShut
  {NoStop}%
\bibitem [{\citenamefont {K\'{a}llay}\ and\ \citenamefont
  {Gauss}(2005)}]{Kallay:6}%
  \BibitemOpen
  \bibfield  {author} {\bibinfo {author} {\bibfnamefont {M.}~\bibnamefont
  {K\'{a}llay}}\ and\ \bibinfo {author} {\bibfnamefont {J.}~\bibnamefont
  {Gauss}},\ }\href {\doibase 10.1063/1.2121589} {\bibfield  {journal}
  {\bibinfo  {journal} {J.\ Chem.\ Phys.}\ }\textbf {\bibinfo {volume} {123}},\
  \bibinfo {eid} {214105} (\bibinfo {year} {2005})}\BibitemShut {NoStop}%
\bibitem [{\citenamefont {Skripnikov}(2021)}]{Skripnikov:2021a}%
  \BibitemOpen
  \bibfield  {author} {\bibinfo {author} {\bibfnamefont {L.~V.}\ \bibnamefont
  {Skripnikov}},\ }\href {\doibase 10.1063/5.0053659} {\bibfield  {journal}
  {\bibinfo  {journal} {J.\ Chem.\ Phys.}\ }\textbf {\bibinfo {volume} {154}},\
  \bibinfo {pages} {201101} (\bibinfo {year} {2021})}\BibitemShut {NoStop}%
\bibitem [{\citenamefont {Athanasakis-Kaklamanakis}\ \emph
  {et~al.}(2023)\citenamefont {Athanasakis-Kaklamanakis}, \citenamefont
  {Wilkins}, \citenamefont {Skripnikov}, \citenamefont {Koszorus},
  \citenamefont {Breier}, \citenamefont {Au}, \citenamefont {Belosevic},
  \citenamefont {Berger}, \citenamefont {Bissell}, \citenamefont {Borschevsky},
  \citenamefont {Brinson}, \citenamefont {Chrysalidis}, \citenamefont
  {Cocolios}, \citenamefont {de~Groote}, \citenamefont {Dorne}, \citenamefont
  {Fajardo-Zambrano}, \citenamefont {Field}, \citenamefont {Flanagan},
  \citenamefont {Franchoo}, \citenamefont {Ruiz}, \citenamefont {Gaul},
  \citenamefont {Geldhof}, \citenamefont {Giesen}, \citenamefont {Hanstorp},
  \citenamefont {Heinke}, \citenamefont {Isaev}, \citenamefont {Kyuberis},
  \citenamefont {Kujanpaa}, \citenamefont {Lalanne}, \citenamefont {Neyens},
  \citenamefont {Nichols}, \citenamefont {Pasteka}, \citenamefont {Perrett},
  \citenamefont {Reilly}, \citenamefont {Rothe}, \citenamefont {Udrescu},
  \citenamefont {van~den Borne}, \citenamefont {Wang}, \citenamefont
  {Wessolek}, \citenamefont {Yang},\ and\ \citenamefont
  {Zuelch}}]{AthanasakisRaFPinning:2023}%
  \BibitemOpen
  \bibfield  {author} {\bibinfo {author} {\bibfnamefont {M.}~\bibnamefont
  {Athanasakis-Kaklamanakis}}, \bibinfo {author} {\bibfnamefont {S.~G.}\
  \bibnamefont {Wilkins}}, \bibinfo {author} {\bibfnamefont {L.~V.}\
  \bibnamefont {Skripnikov}}, \bibinfo {author} {\bibfnamefont
  {A.}~\bibnamefont {Koszorus}}, \bibinfo {author} {\bibfnamefont {A.~A.}\
  \bibnamefont {Breier}}, \bibinfo {author} {\bibfnamefont {M.}~\bibnamefont
  {Au}}, \bibinfo {author} {\bibfnamefont {I.}~\bibnamefont {Belosevic}},
  \bibinfo {author} {\bibfnamefont {R.}~\bibnamefont {Berger}}, \bibinfo
  {author} {\bibfnamefont {M.~L.}\ \bibnamefont {Bissell}}, \bibinfo {author}
  {\bibfnamefont {A.}~\bibnamefont {Borschevsky}}, \bibinfo {author}
  {\bibfnamefont {A.}~\bibnamefont {Brinson}}, \bibinfo {author} {\bibfnamefont
  {K.}~\bibnamefont {Chrysalidis}}, \bibinfo {author} {\bibfnamefont {T.~E.}\
  \bibnamefont {Cocolios}}, \bibinfo {author} {\bibfnamefont {R.~P.}\
  \bibnamefont {de~Groote}}, \bibinfo {author} {\bibfnamefont {A.}~\bibnamefont
  {Dorne}}, \bibinfo {author} {\bibfnamefont {C.~M.}\ \bibnamefont
  {Fajardo-Zambrano}}, \bibinfo {author} {\bibfnamefont {R.~W.}\ \bibnamefont
  {Field}}, \bibinfo {author} {\bibfnamefont {K.~T.}\ \bibnamefont {Flanagan}},
  \bibinfo {author} {\bibfnamefont {S.}~\bibnamefont {Franchoo}}, \bibinfo
  {author} {\bibfnamefont {R.~F.~G.}\ \bibnamefont {Ruiz}}, \bibinfo {author}
  {\bibfnamefont {K.}~\bibnamefont {Gaul}}, \bibinfo {author} {\bibfnamefont
  {S.}~\bibnamefont {Geldhof}}, \bibinfo {author} {\bibfnamefont {T.~F.}\
  \bibnamefont {Giesen}}, \bibinfo {author} {\bibfnamefont {D.}~\bibnamefont
  {Hanstorp}}, \bibinfo {author} {\bibfnamefont {R.}~\bibnamefont {Heinke}},
  \bibinfo {author} {\bibfnamefont {T.~A.}\ \bibnamefont {Isaev}}, \bibinfo
  {author} {\bibfnamefont {A.~A.}\ \bibnamefont {Kyuberis}}, \bibinfo {author}
  {\bibfnamefont {S.}~\bibnamefont {Kujanpaa}}, \bibinfo {author}
  {\bibfnamefont {L.}~\bibnamefont {Lalanne}}, \bibinfo {author} {\bibfnamefont
  {G.}~\bibnamefont {Neyens}}, \bibinfo {author} {\bibfnamefont
  {M.}~\bibnamefont {Nichols}}, \bibinfo {author} {\bibfnamefont {L.~F.}\
  \bibnamefont {Pasteka}}, \bibinfo {author} {\bibfnamefont {H.~A.}\
  \bibnamefont {Perrett}}, \bibinfo {author} {\bibfnamefont {J.~R.}\
  \bibnamefont {Reilly}}, \bibinfo {author} {\bibfnamefont {S.}~\bibnamefont
  {Rothe}}, \bibinfo {author} {\bibfnamefont {S.~M.}\ \bibnamefont {Udrescu}},
  \bibinfo {author} {\bibfnamefont {B.}~\bibnamefont {van~den Borne}}, \bibinfo
  {author} {\bibfnamefont {Q.}~\bibnamefont {Wang}}, \bibinfo {author}
  {\bibfnamefont {J.}~\bibnamefont {Wessolek}}, \bibinfo {author}
  {\bibfnamefont {X.~F.}\ \bibnamefont {Yang}}, \ and\ \bibinfo {author}
  {\bibfnamefont {C.}~\bibnamefont {Zuelch}},\ }\href@noop {} {\enquote
  {\bibinfo {title} {Pinning down electron correlations in raf via spectroscopy
  of excited states},}\ } (\bibinfo {year} {2023}),\ \Eprint
  {http://arxiv.org/abs/2308.14862} {arXiv:2308.14862 [physics.atom-ph]}
  \BibitemShut {NoStop}%
\bibitem [{\citenamefont {Sikkema}\ \emph {et~al.}(2009)\citenamefont
  {Sikkema}, \citenamefont {Visscher}, \citenamefont {Saue},\ and\
  \citenamefont {Ilia\v{s}}}]{Sikkema:2009}%
  \BibitemOpen
  \bibfield  {author} {\bibinfo {author} {\bibfnamefont {J.}~\bibnamefont
  {Sikkema}}, \bibinfo {author} {\bibfnamefont {L.}~\bibnamefont {Visscher}},
  \bibinfo {author} {\bibfnamefont {T.}~\bibnamefont {Saue}}, \ and\ \bibinfo
  {author} {\bibfnamefont {M.}~\bibnamefont {Ilia\v{s}}},\ }\href {\doibase   10.1063/1.3239505} {\bibfield  {journal} {\bibinfo  {journal} {J.\ Chem.\
  Phys.}\ }\textbf {\bibinfo {volume} {131}},\ \bibinfo {pages} {124116}
  (\bibinfo {year} {2009})}\BibitemShut {NoStop}%
\bibitem [{\citenamefont {Shabaev}\ \emph {et~al.}(2013)\citenamefont
  {Shabaev}, \citenamefont {Tupitsyn},\ and\ \citenamefont
  {Yerokhin}}]{Shabaev:13}%
  \BibitemOpen
  \bibfield  {author} {\bibinfo {author} {\bibfnamefont {V.~M.}\ \bibnamefont
  {Shabaev}}, \bibinfo {author} {\bibfnamefont {I.~I.}\ \bibnamefont
  {Tupitsyn}}, \ and\ \bibinfo {author} {\bibfnamefont {V.~A.}\ \bibnamefont
  {Yerokhin}},\ }\href {\doibase 10.1103/PhysRevA.88.012513} {\bibfield
  {journal} {\bibinfo  {journal} {Phys.\ Rev.\ A}\ }\textbf {\bibinfo {volume}
  {88}},\ \bibinfo {pages} {012513} (\bibinfo {year} {2013})}\BibitemShut
  {NoStop}%
\bibitem [{\citenamefont {Malyshev}\ \emph {et~al.}(2022)\citenamefont
  {Malyshev}, \citenamefont {Glazov}, \citenamefont {Shabaev}, \citenamefont
  {Tupitsyn}, \citenamefont {Yerokhin},\ and\ \citenamefont
  {Zaytsev}}]{Malyshev:2022}%
  \BibitemOpen
  \bibfield  {author} {\bibinfo {author} {\bibfnamefont {A.~V.}\ \bibnamefont
  {Malyshev}}, \bibinfo {author} {\bibfnamefont {D.~A.}\ \bibnamefont
  {Glazov}}, \bibinfo {author} {\bibfnamefont {V.~M.}\ \bibnamefont {Shabaev}},
  \bibinfo {author} {\bibfnamefont {I.~I.}\ \bibnamefont {Tupitsyn}}, \bibinfo
  {author} {\bibfnamefont {V.~A.}\ \bibnamefont {Yerokhin}}, \ and\ \bibinfo
  {author} {\bibfnamefont {V.~A.}\ \bibnamefont {Zaytsev}},\ }\href {\doibase   10.1103/PhysRevA.106.012806} {\bibfield  {journal} {\bibinfo  {journal}
  {Phys. Rev. A}\ }\textbf {\bibinfo {volume} {106}},\ \bibinfo {pages}
  {012806} (\bibinfo {year} {2022})}\BibitemShut {NoStop}%
\bibitem [{\citenamefont {Maison}\ \emph {et~al.}(2019)\citenamefont {Maison},
  \citenamefont {Skripnikov},\ and\ \citenamefont {Glazov}}]{Maison:2019}%
  \BibitemOpen
  \bibfield  {author} {\bibinfo {author} {\bibfnamefont {D.~E.}\ \bibnamefont
  {Maison}}, \bibinfo {author} {\bibfnamefont {L.~V.}\ \bibnamefont
  {Skripnikov}}, \ and\ \bibinfo {author} {\bibfnamefont {D.~A.}\ \bibnamefont
  {Glazov}},\ }\href {\doibase 10.1103/PhysRevA.99.042506} {\bibfield
  {journal} {\bibinfo  {journal} {Phys. Rev. A}\ }\textbf {\bibinfo {volume}
  {99}},\ \bibinfo {pages} {042506} (\bibinfo {year} {2019})}\BibitemShut
  {NoStop}%
\bibitem [{\citenamefont {Johnson}\ and\ \citenamefont
  {Soff}(1985)}]{Johnson:1985}%
  \BibitemOpen
  \bibfield  {author} {\bibinfo {author} {\bibfnamefont {W.}~\bibnamefont
  {Johnson}}\ and\ \bibinfo {author} {\bibfnamefont {G.}~\bibnamefont {Soff}},\
  }\href {\doibase https://doi.org/10.1016/0092-640X(85)90010-5} {\bibfield
  {journal} {\bibinfo  {journal} {At. Data Nucl. Data Tables}\ }\textbf
  {\bibinfo {volume} {33}},\ \bibinfo {pages} {405} (\bibinfo {year}
  {1985})}\BibitemShut {NoStop}%
\bibitem [{\citenamefont {Visscher}\ and\ \citenamefont
  {Dyall}(1997)}]{Visscher:1997}%
  \BibitemOpen
  \bibfield  {author} {\bibinfo {author} {\bibfnamefont {L.}~\bibnamefont
  {Visscher}}\ and\ \bibinfo {author} {\bibfnamefont {K.~G.}\ \bibnamefont
  {Dyall}},\ }\href {\doibase 10.1006/adnd.1997.0751} {\bibfield  {journal}
  {\bibinfo  {journal} {At. Data Nucl. Data Tables}\ }\textbf {\bibinfo
  {volume} {67}},\ \bibinfo {pages} {207} (\bibinfo {year} {1997})}\BibitemShut
  {NoStop}%
\bibitem [{DIR()}]{DIRAC19}%
  \BibitemOpen
  \href@noop {} {}\bibinfo {note} {DIRAC, a relativistic ab initio electronic
  structure program, Release DIRAC19 (2019), written by A. S. P. Gomes, T.
  Saue, L. Visscher, H. J. Aa. Jensen, and R. Bast, with contributions from I.
  A. Aucar, V. Bakken, K. G. Dyall, S. Dubillard, U. Ekstroem, E. Eliav, T.
  Enevoldsen, E. Fasshauer, T. Fleig, O. Fossgaard, L. Halbert, E. D.
  Hedegaard, T. Helgaker, J. Henriksson, M. Ilias, Ch. R. Jacob, S. Knecht, S.
  Komorovsky, O. Kullie, J. K. Laerdahl, C. V. Larsen, Y. S. Lee, H. S.
  Nataraj, M. K. Nayak, P. Norman, M. Olejniczak, J. Olsen, J. M. H. Olsen, Y.
  C. Park, J. K. Pedersen, M. Pernpointner, R. Di Remigio, K. Ruud, P. Salek,
  B. Schimmelpfennig, B. Senjean, A. Shee, J. Sikkema, A. J. Thorvaldsen, J.
  Thyssen, J. van Stralen, M. L. Vidal, S. Villaume, O. Visser, T. Winther, and
  S. Yamamoto (see http://diracprogram.org).}\BibitemShut {Stop}%
\bibitem [{\citenamefont {Saue}\ \emph {et~al.}(2020)\citenamefont {Saue},
  \citenamefont {Bast}, \citenamefont {Gomes}, \citenamefont {Jensen},
  \citenamefont {Visscher}, \citenamefont {Aucar}, \citenamefont {Di~Remigio},
  \citenamefont {Dyall}, \citenamefont {Eliav}, \citenamefont {Fasshauer},
  \citenamefont {Fleig}, \citenamefont {Halbert}, \citenamefont {Hedegard},
  \citenamefont {Helmich-Paris}, \citenamefont {Ilias}, \citenamefont {Jacob},
  \citenamefont {Knecht}, \citenamefont {Laerdahl}, \citenamefont {Vidal},
  \citenamefont {Nayak}, \citenamefont {Olejniczak}, \citenamefont {Olsen},
  \citenamefont {Pernpointner}, \citenamefont {Senjean}, \citenamefont {Shee},
  \citenamefont {Sunaga},\ and\ \citenamefont {van Stralen}}]{Saue:2020}%
  \BibitemOpen
  \bibfield  {author} {\bibinfo {author} {\bibfnamefont {T.}~\bibnamefont
  {Saue}}, \bibinfo {author} {\bibfnamefont {R.}~\bibnamefont {Bast}}, \bibinfo
  {author} {\bibfnamefont {A.~S.~P.}\ \bibnamefont {Gomes}}, \bibinfo {author}
  {\bibfnamefont {H.~J.~A.}\ \bibnamefont {Jensen}}, \bibinfo {author}
  {\bibfnamefont {L.}~\bibnamefont {Visscher}}, \bibinfo {author}
  {\bibfnamefont {I.~A.}\ \bibnamefont {Aucar}}, \bibinfo {author}
  {\bibfnamefont {R.}~\bibnamefont {Di~Remigio}}, \bibinfo {author}
  {\bibfnamefont {K.~G.}\ \bibnamefont {Dyall}}, \bibinfo {author}
  {\bibfnamefont {E.}~\bibnamefont {Eliav}}, \bibinfo {author} {\bibfnamefont
  {E.}~\bibnamefont {Fasshauer}}, \bibinfo {author} {\bibfnamefont
  {T.}~\bibnamefont {Fleig}}, \bibinfo {author} {\bibfnamefont
  {L.}~\bibnamefont {Halbert}}, \bibinfo {author} {\bibfnamefont {E.~D.}\
  \bibnamefont {Hedegard}}, \bibinfo {author} {\bibfnamefont {B.}~\bibnamefont
  {Helmich-Paris}}, \bibinfo {author} {\bibfnamefont {M.}~\bibnamefont
  {Ilias}}, \bibinfo {author} {\bibfnamefont {C.~R.}\ \bibnamefont {Jacob}},
  \bibinfo {author} {\bibfnamefont {S.}~\bibnamefont {Knecht}}, \bibinfo
  {author} {\bibfnamefont {J.~K.}\ \bibnamefont {Laerdahl}}, \bibinfo {author}
  {\bibfnamefont {M.~L.}\ \bibnamefont {Vidal}}, \bibinfo {author}
  {\bibfnamefont {M.~K.}\ \bibnamefont {Nayak}}, \bibinfo {author}
  {\bibfnamefont {M.}~\bibnamefont {Olejniczak}}, \bibinfo {author}
  {\bibfnamefont {J.~M.~H.}\ \bibnamefont {Olsen}}, \bibinfo {author}
  {\bibfnamefont {M.}~\bibnamefont {Pernpointner}}, \bibinfo {author}
  {\bibfnamefont {B.}~\bibnamefont {Senjean}}, \bibinfo {author} {\bibfnamefont
  {A.}~\bibnamefont {Shee}}, \bibinfo {author} {\bibfnamefont {A.}~\bibnamefont
  {Sunaga}}, \ and\ \bibinfo {author} {\bibfnamefont {J.~N.~P.}\ \bibnamefont
  {van Stralen}},\ }\href {\doibase 10.1063/5.0004844} {\bibfield  {journal}
  {\bibinfo  {journal} {J.\ Chem.\ Phys.}\ }\textbf {\bibinfo {volume} {152}},\
  \bibinfo {pages} {204104} (\bibinfo {year} {2020})}\BibitemShut {NoStop}%
\bibitem [{MRC()}]{MRCC2020}%
  \BibitemOpen
  \href@noop {} {\enquote {\bibinfo {title} {{{\sc mrcc}}},}\ }\bibinfo {note}
  {M. K\'{a}llay, P. R. Nagy, D. Mester, Z. Rolik, G. Samu, J. Csontos, J.
  Cs\'{o}ka, P. B. Szab\'{o}, L. Gyevi-Nagy, B. H\'{e}gely, I. Ladj\'{a}nszki,
  L. Szegedy, B. Lad\'{o}czki, K. Petrov, M. Farkas, P. D. Mezei, and \'{a}.
  Ganyecz: The {\sc mrcc} program system: Accurate quantum chemistry from water
  to proteins, J. Chem. Phys. 152, 074107 (2020); {\sc mrcc}, a quantum
  chemical program suite written by M. K\'{a}llay, P. R. Nagy, D. Mester, Z.
  Rolik, G. Samu, J. Csontos, J. Cs\'{o}ka, P. B. Szab\'{o}, L. Gyevi-Nagy, B.
  H\'{e}gely, I. Ladj\'{a}nszki, L. Szegedy, B. Lad\'{o}czki, K. Petrov, M.
  Farkas, P. D. Mezei, and \'{A}. Ganyecz. See www.mrcc.hu.}\BibitemShut
  {Stop}%
\bibitem [{\citenamefont {K\'{a}llay}\ and\ \citenamefont
  {Surj\'{a}n}(2001)}]{Kallay:1}%
  \BibitemOpen
  \bibfield  {author} {\bibinfo {author} {\bibfnamefont {M.}~\bibnamefont
  {K\'{a}llay}}\ and\ \bibinfo {author} {\bibfnamefont {P.~R.}\ \bibnamefont
  {Surj\'{a}n}},\ }\href {\doibase 10.1063/1.1383290} {\bibfield  {journal}
  {\bibinfo  {journal} {J.\ Chem.\ Phys.}\ }\textbf {\bibinfo {volume} {115}},\
  \bibinfo {pages} {2945} (\bibinfo {year} {2001})}\BibitemShut {NoStop}%
\bibitem [{\citenamefont {K\'{a}llay}\ \emph {et~al.}(2002)\citenamefont
  {K\'{a}llay}, \citenamefont {Szalay},\ and\ \citenamefont
  {Surj\'{a}n}}]{Kallay:2}%
  \BibitemOpen
  \bibfield  {author} {\bibinfo {author} {\bibfnamefont {M.}~\bibnamefont
  {K\'{a}llay}}, \bibinfo {author} {\bibfnamefont {P.~G.}\ \bibnamefont
  {Szalay}}, \ and\ \bibinfo {author} {\bibfnamefont {P.~R.}\ \bibnamefont
  {Surj\'{a}n}},\ }\href {\doibase 10.1063/1.1483856} {\bibfield  {journal}
  {\bibinfo  {journal} {J.\ Chem.\ Phys.}\ }\textbf {\bibinfo {volume} {117}},\
  \bibinfo {pages} {980} (\bibinfo {year} {2002})}\BibitemShut {NoStop}%
\bibitem [{\citenamefont {Oleynichenko}\ \emph {et~al.}(2021)\citenamefont
  {Oleynichenko}, \citenamefont {Zaitsevskii},\ and\ \citenamefont
  {Eliav}}]{EXPT_website}%
  \BibitemOpen
  \bibfield  {author} {\bibinfo {author} {\bibfnamefont {A.}~\bibnamefont
  {Oleynichenko}}, \bibinfo {author} {\bibfnamefont {A.}~\bibnamefont
  {Zaitsevskii}}, \ and\ \bibinfo {author} {\bibfnamefont {E.}~\bibnamefont
  {Eliav}},\ }\href@noop {} {} (\bibinfo {year} {2021}),\ \bibinfo {note}
  {{EXP-T}, an extensible code for {F}ock space relativistic coupled cluster
  calculations (see \url{http://www.qchem.pnpi.spb.ru/expt})}\BibitemShut
  {NoStop}%
\bibitem [{\citenamefont {Oleynichenko}\ \emph {et~al.}(2020)\citenamefont
  {Oleynichenko}, \citenamefont {Zaitsevskii},\ and\ \citenamefont
  {Eliav}}]{Oleynichenko_EXPT}%
  \BibitemOpen
  \bibfield  {author} {\bibinfo {author} {\bibfnamefont {A.~V.}\ \bibnamefont
  {Oleynichenko}}, \bibinfo {author} {\bibfnamefont {A.}~\bibnamefont
  {Zaitsevskii}}, \ and\ \bibinfo {author} {\bibfnamefont {E.}~\bibnamefont
  {Eliav}},\ }in\ \href {\doibase 10.1007/978-3-030-64616-5_33} {\emph
  {\bibinfo {booktitle} {Supercomputing}}},\ Vol.\ \bibinfo {volume} {1331},\
  \bibinfo {editor} {edited by\ \bibinfo {editor} {\bibfnamefont
  {V.}~\bibnamefont {Voevodin}}\ and\ \bibinfo {editor} {\bibfnamefont
  {S.}~\bibnamefont {Sobolev}}}\ (\bibinfo  {publisher} {Springer International
  Publishing},\ \bibinfo {address} {Cham},\ \bibinfo {year} {2020})\ pp.\
  \bibinfo {pages} {375--386}\BibitemShut {NoStop}%
\bibitem [{\citenamefont {A.~Zaitsevskii}\ and\ \citenamefont
  {Eliav}(2023)}]{Zaitsevskii:2023}%
  \BibitemOpen
  \bibfield  {author} {\bibinfo {author} {\bibfnamefont {A.~V.~O.}\
  \bibnamefont {A.~Zaitsevskii}}\ and\ \bibinfo {author} {\bibfnamefont
  {E.}~\bibnamefont {Eliav}},\ }\href {\doibase 10.1080/00268976.2023.2236246}
  {\bibfield  {journal} {\bibinfo  {journal} {Mol. Phys.}\ }\textbf {\bibinfo
  {volume} {0}},\ \bibinfo {pages} {e2236246} (\bibinfo {year}
  {2023})}\BibitemShut {NoStop}%
\bibitem [{\citenamefont {Stanton}\ \emph {et~al.}(2011)\citenamefont
  {Stanton}, \citenamefont {Gauss}, \citenamefont {Harding}, \citenamefont
  {Szalay} \emph {et~al.}}]{CFOUR}%
  \BibitemOpen
  \bibfield  {author} {\bibinfo {author} {\bibfnamefont {J.~F.}\ \bibnamefont
  {Stanton}}, \bibinfo {author} {\bibfnamefont {J.}~\bibnamefont {Gauss}},
  \bibinfo {author} {\bibfnamefont {M.~E.}\ \bibnamefont {Harding}}, \bibinfo
  {author} {\bibfnamefont {P.~G.}\ \bibnamefont {Szalay}},  \emph {et~al.},\
  }\href@noop {} {\enquote {\bibinfo {title} {{``{\sc cfour}''}},}\ } (\bibinfo
  {year} {2011}),\ \bibinfo {note} {{\sc cfour}: a program package for
  performing high-level quantum chemical calculations on atoms and molecules,
  {http://www.cfour.de} .}\BibitemShut {Stop}%
\bibitem [{\citenamefont {Matthews}\ \emph {et~al.}(2020)\citenamefont
  {Matthews}, \citenamefont {Cheng}, \citenamefont {Harding}, \citenamefont
  {Lipparini}, \citenamefont {Stopkowicz}, \citenamefont {Jagau}, \citenamefont
  {Szalay}, \citenamefont {Gauss},\ and\ \citenamefont
  {Stanton}}]{Matthews:CFOUR:20}%
  \BibitemOpen
  \bibfield  {author} {\bibinfo {author} {\bibfnamefont {D.~A.}\ \bibnamefont
  {Matthews}}, \bibinfo {author} {\bibfnamefont {L.}~\bibnamefont {Cheng}},
  \bibinfo {author} {\bibfnamefont {M.~E.}\ \bibnamefont {Harding}}, \bibinfo
  {author} {\bibfnamefont {F.}~\bibnamefont {Lipparini}}, \bibinfo {author}
  {\bibfnamefont {S.}~\bibnamefont {Stopkowicz}}, \bibinfo {author}
  {\bibfnamefont {T.-C.}\ \bibnamefont {Jagau}}, \bibinfo {author}
  {\bibfnamefont {P.~G.}\ \bibnamefont {Szalay}}, \bibinfo {author}
  {\bibfnamefont {J.}~\bibnamefont {Gauss}}, \ and\ \bibinfo {author}
  {\bibfnamefont {J.~F.}\ \bibnamefont {Stanton}},\ }\href {\doibase   10.1063/5.0004837} {\bibfield  {journal} {\bibinfo  {journal} {J. Chem.
  Phys.}\ }\textbf {\bibinfo {volume} {152}},\ \bibinfo {pages} {214108}
  (\bibinfo {year} {2020})}\BibitemShut {NoStop}%
\bibitem [{\citenamefont {Skripnikov}(2016)}]{Skripnikov:16b}%
  \BibitemOpen
  \bibfield  {author} {\bibinfo {author} {\bibfnamefont {L.~V.}\ \bibnamefont
  {Skripnikov}},\ }\href {\doibase 10.1063/1.4968229} {\bibfield  {journal}
  {\bibinfo  {journal} {J.\ Chem.\ Phys.}\ }\textbf {\bibinfo {volume} {145}},\
  \bibinfo {pages} {214301} (\bibinfo {year} {2016})}\BibitemShut {NoStop}%
\bibitem [{\citenamefont {Peterson}\ \emph {et~al.}(2003)\citenamefont
  {Peterson}, \citenamefont {Figgen}, \citenamefont {Goll}, \citenamefont
  {Stoll},\ and\ \citenamefont {Dolg}}]{Peterson:2003}%
  \BibitemOpen
  \bibfield  {author} {\bibinfo {author} {\bibfnamefont {K.~A.}\ \bibnamefont
  {Peterson}}, \bibinfo {author} {\bibfnamefont {D.}~\bibnamefont {Figgen}},
  \bibinfo {author} {\bibfnamefont {E.}~\bibnamefont {Goll}}, \bibinfo {author}
  {\bibfnamefont {H.}~\bibnamefont {Stoll}}, \ and\ \bibinfo {author}
  {\bibfnamefont {M.}~\bibnamefont {Dolg}},\ }\href {\doibase   10.1063/1.1622924} {\bibfield  {journal} {\bibinfo  {journal} {J. Chem.
  Phys.}\ }\textbf {\bibinfo {volume} {119}},\ \bibinfo {pages} {11113}
  (\bibinfo {year} {2003})}\BibitemShut {NoStop}%
\bibitem [{\citenamefont {Roos}\ \emph {et~al.}(2004)\citenamefont {Roos},
  \citenamefont {Lindh}, \citenamefont {Malmqvist}, \citenamefont {Veryazov},\
  and\ \citenamefont {Widmark}}]{Roos:2004}%
  \BibitemOpen
  \bibfield  {author} {\bibinfo {author} {\bibfnamefont {B.~O.}\ \bibnamefont
  {Roos}}, \bibinfo {author} {\bibfnamefont {R.}~\bibnamefont {Lindh}},
  \bibinfo {author} {\bibfnamefont {P.-A.~k.}\ \bibnamefont {Malmqvist}},
  \bibinfo {author} {\bibfnamefont {V.}~\bibnamefont {Veryazov}}, \ and\
  \bibinfo {author} {\bibfnamefont {P.-O.}\ \bibnamefont {Widmark}},\ }\href
  {\doibase 10.1021/jp031064+} {\bibfield  {journal} {\bibinfo  {journal} {J.
  Phys. Chem. A}\ }\textbf {\bibinfo {volume} {108}},\ \bibinfo {pages} {2851}
  (\bibinfo {year} {2004})}\BibitemShut {NoStop}%
\bibitem [{\citenamefont {Zeng}\ \emph {et~al.}(2010)\citenamefont {Zeng},
  \citenamefont {Fedorov},\ and\ \citenamefont {Klobukowski}}]{Zeng:2010}%
  \BibitemOpen
  \bibfield  {author} {\bibinfo {author} {\bibfnamefont {T.}~\bibnamefont
  {Zeng}}, \bibinfo {author} {\bibfnamefont {D.~G.}\ \bibnamefont {Fedorov}}, \
  and\ \bibinfo {author} {\bibfnamefont {M.}~\bibnamefont {Klobukowski}},\
  }\href {\doibase 10.1063/1.3297887} {\bibfield  {journal} {\bibinfo
  {journal} {J. Chem. Phys.}\ }\textbf {\bibinfo {volume} {132}} (\bibinfo
  {year} {2010}),\ 10.1063/1.3297887},\ \bibinfo {note} {074102}\BibitemShut
  {NoStop}%
\bibitem [{\citenamefont {Laury}\ and\ \citenamefont
  {Wilson}(2012)}]{Wilson:2012}%
  \BibitemOpen
  \bibfield  {author} {\bibinfo {author} {\bibfnamefont {M.~L.}\ \bibnamefont
  {Laury}}\ and\ \bibinfo {author} {\bibfnamefont {A.~K.}\ \bibnamefont
  {Wilson}},\ }\href {\doibase 10.1063/1.4768420} {\bibfield  {journal}
  {\bibinfo  {journal} {J. Chem. Phys.}\ }\textbf {\bibinfo {volume} {137}}
  (\bibinfo {year} {2012}),\ 10.1063/1.4768420}\BibitemShut {NoStop}%
\bibitem [{\citenamefont {Borschevsky}\ \emph {et~al.}(2015)\citenamefont
  {Borschevsky}, \citenamefont {Pa\ifmmode~\check{s}\else \v{s}\fi{}teka},
  \citenamefont {Pershina}, \citenamefont {Eliav},\ and\ \citenamefont
  {Kaldor}}]{Borschevsky:2015}%
  \BibitemOpen
  \bibfield  {author} {\bibinfo {author} {\bibfnamefont {A.}~\bibnamefont
  {Borschevsky}}, \bibinfo {author} {\bibfnamefont {L.~F.}\ \bibnamefont
  {Pa\ifmmode~\check{s}\else \v{s}\fi{}teka}}, \bibinfo {author} {\bibfnamefont
  {V.}~\bibnamefont {Pershina}}, \bibinfo {author} {\bibfnamefont
  {E.}~\bibnamefont {Eliav}}, \ and\ \bibinfo {author} {\bibfnamefont
  {U.}~\bibnamefont {Kaldor}},\ }\href {\doibase 10.1103/PhysRevA.91.020501}
  {\bibfield  {journal} {\bibinfo  {journal} {Phys. Rev. A}\ }\textbf {\bibinfo
  {volume} {91}},\ \bibinfo {pages} {020501} (\bibinfo {year}
  {2015})}\BibitemShut {NoStop}%
\bibitem [{\citenamefont {Skripnikov}\ \emph {et~al.}(2021)\citenamefont
  {Skripnikov}, \citenamefont {Chubukov},\ and\ \citenamefont
  {Shakhova}}]{Skripnikov:2021b}%
  \BibitemOpen
  \bibfield  {author} {\bibinfo {author} {\bibfnamefont {L.~V.}\ \bibnamefont
  {Skripnikov}}, \bibinfo {author} {\bibfnamefont {D.~V.}\ \bibnamefont
  {Chubukov}}, \ and\ \bibinfo {author} {\bibfnamefont {V.~M.}\ \bibnamefont
  {Shakhova}},\ }\href {\doibase 10.1063/5.0068267} {\bibfield  {journal}
  {\bibinfo  {journal} {J.\ Chem.\ Phys.}\ }\textbf {\bibinfo {volume} {155}},\
  \bibinfo {pages} {144103} (\bibinfo {year} {2021})}\BibitemShut {NoStop}%
\bibitem [{\citenamefont {Shabaev}\ \emph {et~al.}(2020)\citenamefont
  {Shabaev}, \citenamefont {Tupitsyn}, \citenamefont {Kaygorodov},
  \citenamefont {Kozhedub}, \citenamefont {Malyshev},\ and\ \citenamefont
  {Mironova}}]{ShabaevMolQED:2020}%
  \BibitemOpen
  \bibfield  {author} {\bibinfo {author} {\bibfnamefont {V.~M.}\ \bibnamefont
  {Shabaev}}, \bibinfo {author} {\bibfnamefont {I.~I.}\ \bibnamefont
  {Tupitsyn}}, \bibinfo {author} {\bibfnamefont {M.~Y.}\ \bibnamefont
  {Kaygorodov}}, \bibinfo {author} {\bibfnamefont {Y.~S.}\ \bibnamefont
  {Kozhedub}}, \bibinfo {author} {\bibfnamefont {A.~V.}\ \bibnamefont
  {Malyshev}}, \ and\ \bibinfo {author} {\bibfnamefont {D.~V.}\ \bibnamefont
  {Mironova}},\ }\href {\doibase 10.1103/PhysRevA.101.052502} {\bibfield
  {journal} {\bibinfo  {journal} {Phys. Rev. A}\ }\textbf {\bibinfo {volume}
  {101}},\ \bibinfo {pages} {052502} (\bibinfo {year} {2020})}\BibitemShut
  {NoStop}%
\bibitem [{\citenamefont {M{\aa}rtensson-Pendrill}\ \emph
  {et~al.}(2003)\citenamefont {M{\aa}rtensson-Pendrill}, \citenamefont
  {Gustavsson},\ and\ \citenamefont {Wilson}}]{maartensson2003atomic}%
  \BibitemOpen
  \bibfield  {author} {\bibinfo {author} {\bibfnamefont {A.}~\bibnamefont
  {M{\aa}rtensson-Pendrill}}, \bibinfo {author} {\bibfnamefont
  {M.}~\bibnamefont {Gustavsson}}, \ and\ \bibinfo {author} {\bibfnamefont
  {S.}~\bibnamefont {Wilson}},\ }\href@noop {} {\bibfield  {journal} {\bibinfo
  {journal} {Handbook of Molecular Physics and Quantum Chemistry}\ }\textbf
  {\bibinfo {volume} {1}},\ \bibinfo {pages} {477} (\bibinfo {year}
  {2003})}\BibitemShut {NoStop}%
\bibitem [{\citenamefont {Cocolios}\ \emph {et~al.}(2011)\citenamefont
  {Cocolios}, \citenamefont {Dexters}, \citenamefont {Seliverstov},
  \citenamefont {Andreyev}, \citenamefont {Antalic}, \citenamefont {Barzakh},
  \citenamefont {Bastin}, \citenamefont {B\"uscher}, \citenamefont {Darby},
  \citenamefont {Fedorov}, \citenamefont {Fedosseyev}, \citenamefont
  {Flanagan}, \citenamefont {Franchoo}, \citenamefont {Fritzsche},
  \citenamefont {Huber}, \citenamefont {Huyse}, \citenamefont {Keupers},
  \citenamefont {K\"oster}, \citenamefont {Kudryavtsev}, \citenamefont
  {Man\'e}, \citenamefont {Marsh}, \citenamefont {Molkanov}, \citenamefont
  {Page}, \citenamefont {Sjoedin}, \citenamefont {Stefan}, \citenamefont
  {Van~de Walle}, \citenamefont {Van~Duppen}, \citenamefont {Venhart},
  \citenamefont {Zemlyanoy}, \citenamefont {Bender},\ and\ \citenamefont
  {Heenen}}]{Cocolios:2011}%
  \BibitemOpen
  \bibfield  {author} {\bibinfo {author} {\bibfnamefont {T.~E.}\ \bibnamefont
  {Cocolios}}, \bibinfo {author} {\bibfnamefont {W.}~\bibnamefont {Dexters}},
  \bibinfo {author} {\bibfnamefont {M.~D.}\ \bibnamefont {Seliverstov}},
  \bibinfo {author} {\bibfnamefont {A.~N.}\ \bibnamefont {Andreyev}}, \bibinfo
  {author} {\bibfnamefont {S.}~\bibnamefont {Antalic}}, \bibinfo {author}
  {\bibfnamefont {A.~E.}\ \bibnamefont {Barzakh}}, \bibinfo {author}
  {\bibfnamefont {B.}~\bibnamefont {Bastin}}, \bibinfo {author} {\bibfnamefont
  {J.}~\bibnamefont {B\"uscher}}, \bibinfo {author} {\bibfnamefont {I.~G.}\
  \bibnamefont {Darby}}, \bibinfo {author} {\bibfnamefont {D.~V.}\ \bibnamefont
  {Fedorov}}, \bibinfo {author} {\bibfnamefont {V.~N.}\ \bibnamefont
  {Fedosseyev}}, \bibinfo {author} {\bibfnamefont {K.~T.}\ \bibnamefont
  {Flanagan}}, \bibinfo {author} {\bibfnamefont {S.}~\bibnamefont {Franchoo}},
  \bibinfo {author} {\bibfnamefont {S.}~\bibnamefont {Fritzsche}}, \bibinfo
  {author} {\bibfnamefont {G.}~\bibnamefont {Huber}}, \bibinfo {author}
  {\bibfnamefont {M.}~\bibnamefont {Huyse}}, \bibinfo {author} {\bibfnamefont
  {M.}~\bibnamefont {Keupers}}, \bibinfo {author} {\bibfnamefont
  {U.}~\bibnamefont {K\"oster}}, \bibinfo {author} {\bibfnamefont
  {Y.}~\bibnamefont {Kudryavtsev}}, \bibinfo {author} {\bibfnamefont
  {E.}~\bibnamefont {Man\'e}}, \bibinfo {author} {\bibfnamefont {B.~A.}\
  \bibnamefont {Marsh}}, \bibinfo {author} {\bibfnamefont {P.~L.}\ \bibnamefont
  {Molkanov}}, \bibinfo {author} {\bibfnamefont {R.~D.}\ \bibnamefont {Page}},
  \bibinfo {author} {\bibfnamefont {A.~M.}\ \bibnamefont {Sjoedin}}, \bibinfo
  {author} {\bibfnamefont {I.}~\bibnamefont {Stefan}}, \bibinfo {author}
  {\bibfnamefont {J.}~\bibnamefont {Van~de Walle}}, \bibinfo {author}
  {\bibfnamefont {P.}~\bibnamefont {Van~Duppen}}, \bibinfo {author}
  {\bibfnamefont {M.}~\bibnamefont {Venhart}}, \bibinfo {author} {\bibfnamefont
  {S.~G.}\ \bibnamefont {Zemlyanoy}}, \bibinfo {author} {\bibfnamefont
  {M.}~\bibnamefont {Bender}}, \ and\ \bibinfo {author} {\bibfnamefont {P.-H.}\
  \bibnamefont {Heenen}},\ }\href {\doibase 10.1103/PhysRevLett.106.052503}
  {\bibfield  {journal} {\bibinfo  {journal} {Phys. Rev. Lett.}\ }\textbf
  {\bibinfo {volume} {106}},\ \bibinfo {pages} {052503} (\bibinfo {year}
  {2011})}\BibitemShut {NoStop}%
\bibitem [{\citenamefont {Landau}\ and\ \citenamefont {Lifshitz}(1977)}]{LL77}%
  \BibitemOpen
  \bibfield  {author} {\bibinfo {author} {\bibfnamefont {L.~D.}\ \bibnamefont
  {Landau}}\ and\ \bibinfo {author} {\bibfnamefont {E.~M.}\ \bibnamefont
  {Lifshitz}},\ }\href@noop {} {\emph {\bibinfo {title} {Quantum mechanics}}},\
  \bibinfo {edition} {3rd}\ ed.\ (\bibinfo  {publisher} {Pergamon},\ \bibinfo
  {address} {Oxford},\ \bibinfo {year} {1977})\BibitemShut {NoStop}%
\bibitem [{\citenamefont {Prosnyak}\ \emph {et~al.}(2020)\citenamefont
  {Prosnyak}, \citenamefont {Maison},\ and\ \citenamefont
  {Skripnikov}}]{Prosnyak:2020}%
  \BibitemOpen
  \bibfield  {author} {\bibinfo {author} {\bibfnamefont {S.~D.}\ \bibnamefont
  {Prosnyak}}, \bibinfo {author} {\bibfnamefont {D.~E.}\ \bibnamefont
  {Maison}}, \ and\ \bibinfo {author} {\bibfnamefont {L.~V.}\ \bibnamefont
  {Skripnikov}},\ }\href {\doibase 10.1063/1.5141090} {\bibfield  {journal}
  {\bibinfo  {journal} {J.\ Chem.\ Phys.}\ }\textbf {\bibinfo {volume} {152}},\
  \bibinfo {pages} {044301} (\bibinfo {year} {2020})}\BibitemShut {NoStop}%
\bibitem [{\citenamefont {Shabaev}(1994)}]{shabaev1994hyperfine}%
  \BibitemOpen
  \bibfield  {author} {\bibinfo {author} {\bibfnamefont {V.~M.}\ \bibnamefont
  {Shabaev}},\ }\href {\doibase 10.1088/0953-4075/27/24/006} {\bibfield
  {journal} {\bibinfo  {journal} {J.\ Phys.\ B}\ }\textbf {\bibinfo {volume}
  {27}},\ \bibinfo {pages} {5825} (\bibinfo {year} {1994})}\BibitemShut
  {NoStop}%
\bibitem [{\citenamefont {Reimann}\ and\ \citenamefont
  {McDermott}(1973)}]{Reimann:1973}%
  \BibitemOpen
  \bibfield  {author} {\bibinfo {author} {\bibfnamefont {R.~J.}\ \bibnamefont
  {Reimann}}\ and\ \bibinfo {author} {\bibfnamefont {M.~N.}\ \bibnamefont
  {McDermott}},\ }\href {\doibase 10.1103/PhysRevC.7.2065} {\bibfield
  {journal} {\bibinfo  {journal} {Phys. Rev. C}\ }\textbf {\bibinfo {volume}
  {7}},\ \bibinfo {pages} {2065} (\bibinfo {year} {1973})}\BibitemShut
  {NoStop}%
\bibitem [{\citenamefont {Wouters}\ \emph {et~al.}(1991)\citenamefont
  {Wouters}, \citenamefont {Severijns}, \citenamefont {Vanhaverbeke},\ and\
  \citenamefont {Vanneste}}]{Wouters:1991}%
  \BibitemOpen
  \bibfield  {author} {\bibinfo {author} {\bibfnamefont {J.}~\bibnamefont
  {Wouters}}, \bibinfo {author} {\bibfnamefont {N.}~\bibnamefont {Severijns}},
  \bibinfo {author} {\bibfnamefont {J.}~\bibnamefont {Vanhaverbeke}}, \ and\
  \bibinfo {author} {\bibfnamefont {L.}~\bibnamefont {Vanneste}},\ }\href
  {\doibase 10.1088/0954-3899/17/11/014} {\bibfield  {journal} {\bibinfo
  {journal} {J. Phys. G: Nucl. Part. Phys.}\ }\textbf {\bibinfo {volume}
  {17}},\ \bibinfo {pages} {1673} (\bibinfo {year} {1991})}\BibitemShut
  {NoStop}%
\bibitem [{\citenamefont {Sternheimer}(1950)}]{Sternheimer:1950}%
  \BibitemOpen
  \bibfield  {author} {\bibinfo {author} {\bibfnamefont {R.}~\bibnamefont
  {Sternheimer}},\ }\href {\doibase 10.1103/PhysRev.80.102.2} {\bibfield
  {journal} {\bibinfo  {journal} {Phys. Rev.}\ }\textbf {\bibinfo {volume}
  {80}},\ \bibinfo {pages} {102} (\bibinfo {year} {1950})}\BibitemShut
  {NoStop}%
\end{thebibliography}

%

\end{document}